\documentclass[12pt]{article}
\usepackage{amsmath}
\usepackage{amssymb}
\usepackage{graphics}
\usepackage{times}
\begin{document}

\begin{center}
{\LARGE \textbf{The $\Xi$-transform for conformally flat space-time}\bigskip\ 
\vspace{10pt}\\[0pt]
}

\bigskip {\LARGE \ }

{\large George A.J. Sparling\\Laboratory of Axiomatics\\
Department of Mathematics\\
University of Pittsburgh\\\vspace{3pt} Pittsburgh, Pennsylvania, 15260, USA}

{\vspace{40pt} \large\textbf{Abstract}}\\\vspace{-5pt}
\begin{quote}\textbf{The $\Xi$-transform is a new spinor transform arising naturally in Einstein's general relativity.  Here the example of conformally flat space-time is discussed in detail.  In particular it is shown that for this case, the transform coincides with two other naturally defined transforms: one a two-variable transform on the Lie group $\mathbb{SU}(2, \mathbb{C})$, the other a transform on the space of null split octaves. The key properties of the transform are developed.}\end{quote}
\end{center}
\thispagestyle{empty}
\pagebreak

\section*{Introduction}
The $\Xi$-transform is a transform arising naturally in general relativity, which was found earlier this year by the author as the culmination of a long development \cite{moab13}-\cite{hu1}.  If $\mathbb{M}$ is a space-time (space and time orientable, with a fixed spin structure), denote by $\mathbb{S}^*$ the co-spin bundle of $\mathbb{M}$, the space of all pairs $(x, \pi)$, with $x \in \mathbb{M}$ and $\pi$ a Weyl co-spinor at $x$, so $\mathbb{S}^*$ is a complex vector bundle over $\mathbb{M}$, whose complex fiber dimension is two.  For convenience, we delete the zero section, so $\pi \ne 0$. Denote by $\mathcal{N}$ the horizontal vector field on $\mathbb{S}^*$, that represents the null geodesic spray.  A twistor function $f(x, \pi)$ is by definition a function on $\mathbb{S}^*$  that is killed by $\mathcal{N}$ \cite{pen5}-\cite{pen1}.  The function $f(x, \pi)$ is said to be homogeneous of integral degree $k$ if and only if $f(x, t\pi) = t^k f(x, \pi)$, for any real non-zero $t$.  Denote by $\mathcal{H}_k$ the space of all twistor functions, homogeneous of degree $k$.  Then the $\Xi$-transform is a conformally invariant transform mapping $\mathcal{H}_{-4}$ to $\mathcal{H}_{-2}$.   The basic definition of the transform is given in detail in the author's preprint \cite{moab13}.    In this work we analyze the transform for the case of conformally flat space-time.    We discuss three versions of the transform, here denoted by $\Xi_1$, $\Xi_2$ and $\Xi_3$:
\begin{itemize} \item $\displaystyle{ \Xi_1(f)(g, h) = \int_{p \in \mathbb{G}} f(p, g^{-1}ph) \omega_p}. $\\
Here $\mathbb{G}$ is a compact Lie group, $\omega_p$ is Haar measure, and $f$ is a smooth function of two $\mathbb{G}$-variables.   Then the output $\Xi_1(f)$ is a smooth function of its two arguments.
\item $\displaystyle{ \Xi_2(f)(x, \eta) = i\int_{\pi_{A'} \eta^{A'} = 1}  f(x^a + s\overline{\eta}^A \eta^{A'}, \pi_{C'}) ds \pi^{B'} d\pi_{B'} \overline{\pi}^B d\overline{\pi}_B}.$\\
Here we use two-component spinor notation \cite{pen5}-\cite{pen1}.   The function $f(x^a, \pi_{A'})$ is a twistor function, so is constant along null geodesics:   $\pi^{A'} \overline{\pi}^A \partial_a f(x^b, \pi_{B'}) = 0$ and is homogeneous of degree minus four:  $f(x, t\pi) = t^{-4}$, for $t$ real and non-zero.  Then the output function $\Xi_2(f)$ is also a twistor function, this time homogeneous of degree minus two.   
\end{itemize}
At first sight, these two transformations may  seem unrelated.   We link them by invoking a third transformation, which uses the triality theory of Elie Cartan for spinors associated to the group $\mathbb{O}(4, 4, \mathbb{R})$ \cite{ba1}-\cite{ca1}, \cite{hu1}, \cite{moab11}.  This employs three eight-dimensional real vector spaces, $\mathbb{V}_\alpha$, $\mathbb{V}_\beta$ and $\mathbb{V}_\gamma$,  each equipped with a dot product of neutral signature $(4, 4)$ and linked by the triality map, denoted $(xyz) \in \mathbb{R}$, for any $x$, $y$ and $z$ in  $\mathbb{V}_\alpha$, $\mathbb{V}_\beta$ and $\mathbb{V}_\gamma$.  Dualizing, we obtain product maps, such that $(xyz) = (yz).x = (zx).y = (xy).z$, for any $x$, $y$ and $z$ in  $\mathbb{V}_\alpha$, $\mathbb{V}_\beta$ and $\mathbb{V}_\gamma$, where the products $(yz)$, $(zx)$ and $(xy)$ take values in $\mathbb{V}_\alpha$, $\mathbb{V}_\beta$ and $\mathbb{V}_\gamma$, respectively.
\eject\noindent
The triality obeys the key relations: $((yz)y) = y.y z$, $(z(yz)) = z.z y$, $(x(zx)) = x.x z$, $((zx)z) = z.z x$, $(y(xy)) = y.y x$ and $((xy)x) = x.x y$, for any $x$, $y$ and $z$ in  $\mathbb{V}_\alpha$, $\mathbb{V}_\beta$ and $\mathbb{V}_\gamma$ and the whole theory is invariant under permutations of the three vector spaces.  Such a triality is usually studied in the positive definite case, in which case, if $(xy) = 0$, for $x$ in  $\mathbb{V}_\alpha$ and $y$ in $\mathbb{V}_\beta$, then it follows immediately that either $x$ or $y$ is zero.   Here however, if $(xy) = 0$, we have $0 = ((xy)x) = x.x y $ and $0 = (y(xy)) = y.y x$, from which we see that it is conceivable that $x$ and $y$ can be both  non-zero, provided that they are both null vectors.  Then for $y$  a fixed non-zero null vector in $\mathbb{V}_\beta$, it emerges that the space $\mathcal{N}_y$ of solutions $x \in \mathbb{V}_\alpha$, of the equation $(xy) = 0$ is a four-dimensional completely null self-dual subspace of $\mathbb{V}_\alpha $.   Then the following integral is well-defined:
\begin{itemize} \item $\displaystyle{ \xi_3(f)(y)^{abcd} = \int_{(xy) = 0} f(x) x^{[a} dx^b dx^c dx^{d]}}$.\\
Here $f$ is a smooth function, homogeneous of degree minus four in the variable $x\in \mathbb{V}_\alpha $ and the integral is taken over an oriented  three-sphere representing the fundamental homology class of the complement of the origin in the four-space $\mathcal{N}_y$.   It emerges that $\xi_3(f)(y)^{abcd}$ factorizes: 
\item $\displaystyle{\xi_3(f)(y)^{abcd} =  \Xi_3(f)(y) \sigma^{abcd}(y)}$.\\ Here $\sigma^{abcd} = \sigma^{abcd}_{e'f'} y^{e'}y^{f'}$ and $\sigma^{abcd}_{e'f'} = \sigma^{[abcd]}_{(e'f')}$ and $\sigma^{abcd}_{e'f'}$ is trace-free in its index pair $e'f'$.  Then $\sigma^{[abcd]}_{e'f'}$ gives a natural isomorphism of the space of trace-free symmetric tensors of $\mathbb{V}_\beta$ with the space of self-dual four skew four-index tensors of the space $\mathbb{V}_\alpha$, both spaces being thirty-five dimensional.   Then the coefficient $\Xi_3(f)(y)$ is smooth and homogeneous of degree minus two in the non-zero null vector $y$.   This gives us our third transform.  
\end{itemize}
There are now two main results in this work:
\begin{itemize} \item All three transforms are equivalent, provided we take $\mathbb{G} = \mathbb{SU}(2, \mathbb{C})$ for the case of the transform  $\Xi_1$.  
\item All three transforms obey an equation of the form $\Xi\circ \square = \square \circ \Xi = 0$, where $\square$ is a second-order conformally invariant wave operator (for the ultra-hyperbolic signature $(3, 3)$), whose existence follows from the general theory of  C.  Robin Graham, Ralph Jenne, Lionel Mason and the present author  \cite{gra1}.
\end{itemize} 
We suspect that a stronger result is true: namely that the kernels and images of $\Xi$ and $\square$ exactly match.   However we only prove this here under the technical assumption that the input function has a finite spherical harmonic decomposition.\\\\ 
\begin{itemize}\item    In sections one and two below, we organize the theory of Casimir differential operators on a Lie group. \item  In section three, we describe the spherical harmonics of the group $\mathbb{SU}(2, \mathbb{C})$.  \item  In section four, we reformulate the spherical harmonic theory in terms of two-component spinors. \item  In section five, we introduce the transform and prove its basic properties, for the case of an arbitrary compact Lie group.  In particular for the case when the group is $\mathbb{SU}(2, \mathbb{C})$, we prove the relation $\mathcal{C}\circ \Xi = \Xi\circ \mathcal{C} = 0$, where $\mathcal{C} = C^+ -  C^-$ and $C^+$ and $C^-$ are the standard quadratic Casimir differential operators of $\mathbb{SU}(2, \mathbb{C})$, defined in terms of the first and second variable of the transformed function $\Xi(f)(g, h)$, respectively. \item  In section six, we analyze the kernel and the image of the $\Xi$-transform under the technical assumption that the input function has a finite decomposition in spherical harmonics.  \item In section seven, we introduce the basic null twistor space, which after removing  a real scaling degree of freedom has topology the product of two three-spheres and which is metrically conformal to the metric $G_+ - G_-$, where $G_\pm$ are  unit three-sphere metrics on the two factors of the product.   Geometrically this space represents the space of null geodesics in conformally compactified Minkowksi space-time, where the geodesic is supplied with a tangent co-spinor, defined up to a non-zero real scaling and parallelly propagated along the null geodesic.   Note that the six-dimensional conformal geometry is of the type of Charles Fefferman, associated to the invariant Cauchy-Riemann structure of the null twistor space \cite{moab2}.
\item In section eight, we introduce the basic symmetry groups of the twistor space, the groups $\mathbb{U}(2, 2, \mathbb{C})$ and $\mathbb{O}(4, 4, \mathbb{R})$.
\item In sections nine and ten, we give two different formulations of the null twistor wave operator, $\square$, one using the ambient non-null twistors and the other purely intrinsic.
\item In sections eleven,  twelve and thirteen,  we translate the twistor symmetry generators and the null twistor wave operator into the language of two-component spinors.
\item In section fourteen, we introduce the spinor version of the $\Xi$-transform and then in sections fifteen and sixteen, we prove the basic relations $\Xi\circ \square = \square \circ \Xi = 0$.
\item In section seventeen, we develop the theory of $\mathbb{O}(4, 4)$ triality using quaternion variables and we discuss the incidence relation for null vectors of the triality: the systematic use of quaternion variables tightens up and simplifies the earlier treatment of the author and Philip Tillman \cite{moab11}.
\item In section eighteen, we use twistor variables to parametrize the triality.
\item In section nineteen, we introduce the invariant transform using the twistor variables and prove that the invariant transform encodes the same information as the spinor version of the transform.
\item Finally in section twenty, we reduce the invariant transform to the two-variable group transform, finishing the proof that all three transforms are equivalent.   In particular the group structure of the first transform is understood to arise from a particular choice of conformal factor for the invariant approach.  Thus in the group approach the underlying conformal invariance is not manifest and the relevant functions are not conformally weighted, unlike in the other two cases. 
\end{itemize} 
\eject\noindent
\setcounter{section}{0}
\section{Left and right invariant vector fields}
Let   $\mathbb{G}$ be a compact real Lie group, so $\mathbb{G}$ is also naturally a smooth (even real analytic) compact manifold.    For each $g \in \mathbb{G}$ denote by $L_g: \mathbb{G} \rightarrow \mathbb{G}$ and $R_g: \mathbb{G} \rightarrow \mathbb{G}$ the smooth operations of left multiplication and right multiplication, respectively.   So $L_g(k) = gk$ and $R_g k = kg$, for any $h\in \mathbb{G}$.  Note that $L_g$ and $R_{h}$ commute for any $g$ and $h$ in $\mathbb{G}$.  Also denote by $Ad_g: \mathbb{G} \rightarrow \mathbb{G}$ the adjoint representation of $\mathbb{G}$ on itself, so we have $ Ad_g = L_{g}R_{g^{-1}} $, for any $ g \in \mathbb{G}$.     Denote by $\omega$ the Haar measure for the Lie group, which we represent smoothly by a volume-form $\omega$ with value $\omega_g$ at $g\in \mathbb{G}$.  Then  $\omega$ is invariant under the operations of left multiplication and right multiplication by elements of $\mathbb{G}$, so $\omega$ is also invariant under the adjoint map.
Denote by $\nabla_\alpha$ the Lie algebra of left-invariant vector fields on $\mathbb{G}$.   Here we use Greek indices for the tensors based on the tensor algebra of the tangent and cotangent spaces at the identity of $\mathbb{G}$.   Then we have the Lie bracket:
\[ [\nabla_\alpha, \nabla_\beta] = C_{\alpha\beta}^{\hspace{10pt}\gamma} \nabla_\gamma.\]
Here $C_{\alpha\beta}^{\hspace{10pt}\gamma} = -  C_{\beta\alpha}^{\hspace{10pt}\gamma}$ are the structure constants of the Lie algebra.  
Similarly, denote by $\Delta_\alpha$ the Lie algebra of right-invariant vector fields on $\mathbb{G}$.   Then we have the Lie brackets:
\[ [\nabla_\alpha, \Delta_\beta] = 0, \hspace{10pt} [\Delta_\alpha, \Delta_\beta] = - C_{\alpha\beta}^{\hspace{10pt}\gamma}\Delta_\gamma.\]
Here we have the normalization that at the identity of $\mathbb{G}$, $\Delta_\alpha$ and $\nabla_\alpha$ are equal (and are equal to the Kronecker delta tensor at that point).  Denote by $\textrm{Ad}(g)_\alpha^\beta$ the adjoint action of the Lie group $\mathbb{G}$ on the tangent space to $\mathbb{G}$ at its identity element.    Then we have:
\[ (\Delta_\alpha)_g =  \textrm{Ad}(g^{-1})_\alpha^\beta ( \nabla_\beta)_g.\]
For each vector $v^\alpha$ in the tangent space at the identity of $\mathbb{G}$, denote by $\nabla_v = v^\alpha \nabla_\alpha$ the corresponding left-invariant vector field.   Integrating the vector field $\nabla_v $ gives a one-parameter group of transformations of $\mathbb{G}$, denoted by $R_v(t)$, with $t \in \mathbb{R}$.    We have $R_v(t) = R_{g_v(t)}$, where $\exp_v(t)$ is the integral curve passing through the identity element of the vector field $\nabla_v$.   Also denote by $\Delta_v = v^\alpha \Delta_\alpha $ the corresponding right-invariant vector field.   Integrating the vector field $\Delta_v$ gives a one-parameter group of transformations of $\mathbb{G}$, denoted by $L_v(t)$, with $t \in \mathbb{R}$.    Then we have $L_v(t) = L_{\exp_v(t)}$.   Note that $\exp_v(-t) =  (\exp_v(t))^{-1}$, for any $v$ in the Lie algebra and any real $t$.
\eject\noindent
\section{Casimir operators}
A Casimir operator $C$ for the Lie group $\mathbb{G}$ is, by definition, a differential operator on $\mathbb{G}$, that is both left and right-invariant.   Such an operator may be written as a polynomial in the operators $\nabla_\alpha$ or $\Delta_\alpha$ with constant coefficients.    Modulo the Lie algebra relations, it then has a unique expression as a sum combination of terms $C_n$, $n = 0, 1, 2, \dots$,  where $C_0$ is a constant multiple of the identity and if $n \ge 1$,  $C_n$ takes the following form:
\[ C_n = C^{\alpha_1 \alpha_2 \dots \alpha_n} \nabla_{\alpha_1} \nabla_{\alpha_2} \dots \nabla_{\alpha_n}  = C^{\alpha_1 \alpha_2 \dots \alpha_n} \Delta_{\alpha_1} \Delta_{\alpha_2} \dots \Delta_{\alpha_n} .\]
Here the coefficient tensor $C^{\alpha_1 \alpha_2 \dots \alpha_n}$ is totally symmetric and is both left-invariant and right-invariant.   If $C_n = 0$ for all odd $n$, we say that the Casimir operator is even; if $C_n = 0$, for all even $n$, we say that the Casimir operator is odd.  If only $C_k$ is non-zero, we say that the Casimir operator is of order $k$. \\\\
In the special case that the Lie group is semi-simple, we introduce the Killing form, defined by the formula:
\[ g_{\alpha\beta} = C_{\alpha\gamma}^{\hspace{10pt} \delta} C_{\beta\delta}^{\hspace{10pt} \gamma}.\]
Then it is standard that $g_{\alpha\beta}$ is symmetric and invertible and left and right invariant.   Denoting its inverse by $g^{\alpha\beta}$, we have a preferred second order Casimir operator, denoted by $\square$:
\[ \square = g^{\alpha\beta}  \nabla_{\alpha} \nabla_{\beta} =  g^{\alpha\beta}\Delta_{\alpha} \Delta_{\beta} .\]
In general, if the Lie group has rank $m$, there are $m$ independent Casimir operators, that together with the identity operator generate the algebra of all Casimir operators.  For example in the case of the special unitary group $\mathbb{SU}(n, \mathbb{C})$, consisting of all unitary $n\times n$ complex matrices of unit determinant, we may write the Lie algebra as:
\[ [E^\alpha_\beta, E^\gamma_\delta] = \delta_\beta^\gamma E_\delta^\alpha -  \delta_\delta^\alpha E_\beta^\delta.\]
Here $E^\alpha_\beta$ is trace-free and hermitian.  Then any polynomial in the operator $E^\alpha_\beta$ traced over all its indices, in any fashion, gives a (possibly identically zero) Casimir operator.  The algebra has rank $n-1$, the independent Casimirs being given by the quantities $\textrm{tr}(E^k)$, for $k =2, 3, \dots n$ (where $E^k$ is the $k$-fold matrix product of $E$ with itself).   In particular, for the case $n = 2$, the only independent Casimir operator is the Killing operator, which is proportional to the operator $\textrm{tr}(E^2) = E^\alpha_\beta E^\beta_\alpha$.
\eject\noindent
\section{The case of the Lie group $\mathbb{SU}(2, \mathbb{C})$; harmonics}
The case we need for later is the case of the group $\mathbb{SU}(2, \mathbb{C})$, which we will usually represent as the group of quaternions $q$ that are unit: $q\overline{q} = \overline{q} q = 1$.  If we write $q = t + xi + yj + zk$, where $(t, x, y, z) \in \mathbb{R}^4$, $i^2 = j^2 = k^2 = ijk = -1$, then the group is represented as the unit real three-sphere in $\mathbb{R}^4$, with equation $t^2 + x^2 + y^2 + z^2 = 1$.   A basis for the left invariant vector fields is obtained from the one parameter groups $e^{is}$, $e^{js}$ and $e^{ks}$, where $s$ is real.  The corresponding left-invariant vector fields are as follows:
\[  \underline{J}  =  t \underline{\nabla}  - \underline{x} \partial_t - \underline{x} \times \underline{\nabla}.\]
Here we have $\underline{x}= [x, y, z]$ and $\underline{\nabla} = [\partial_x , \partial_y, \partial_z]$ and we use standard vector notation for Euclidean three-space with co-ordinate vector $\underline{x}$.
Then we have:
\[ \underline{J} \times \underline{J} =  (t \underline{\nabla}  - \underline{x} \partial_t - \underline{x} \times \underline{\nabla}) \times ( t \underline{\nabla}  - \underline{x} \partial_t - \underline{x} \times \underline{\nabla})\]
\[ =  ( \underline{x} \times \underline{\nabla}) t\partial_t +     t (\underline{\nabla} \times ( \underline{\nabla}\times \underline{x}) ) - \underline{x} \partial_t  \times ( t \underline{\nabla})  + \underline{x} \times ( \underline{x} \times \underline{\nabla})\partial_t\]
\[ - (\underline{x} \times \underline{\nabla}) \times ( t \underline{\nabla}  - \underline{x} \partial_t - \underline{x} \times \underline{\nabla})\]
\[ =  - \underline{x} \times \underline{\nabla} +    t \underline{\nabla}(3 + \underline{x}.\underline{\nabla} )  - t (\underline{\nabla}.\underline{\nabla} )\underline{x} - (\underline{x}.\underline{x}) \underline{\nabla} \partial_t  + \underline{x}  (\underline{x}.\underline{\nabla})\partial_t\] \[   + \underline{x} ( \underline{\nabla}. ( t \underline{\nabla}  - \underline{x} \partial_t))  - \underline{\nabla}( \underline{x}. ( t \underline{\nabla}  - \underline{x} \partial_t)) +   t \underline{\nabla}  - \underline{x} \partial_t - \underline{x} \times \underline{\nabla}\]
\[ =   2 t \underline{\nabla}- 2\underline{x} \times \underline{\nabla} - (\underline{x}.\underline{x}) \underline{\nabla} \partial_t   - 4 \underline{x} \partial_t     + \underline{\nabla}( \underline{x}. \underline{x}) \partial_t  = 2\underline{J}.\]
Written out in components, we have:
\[ [J_i, J_j] = 2\epsilon_{ijk} J^k.\]
Here $\epsilon_{ijk}$ is the alternating symbol on three elements, with $\epsilon_{123} = 1$.  The right-invariant vector fields are obtained similarly:
\[  \underline{K}  = t \underline{\nabla}  - \underline{x} \partial_t + \underline{x} \times \underline{\nabla}.\]
Note that at the identity, where $ t = 1$ and $\underline{x} = 0$, the vectors $\underline{J}$ and $\underline{K}$ agree, since they each reduce to just $\underline{\nabla}$.
Then we have:
\[ [K_i, K_j] = - 2\epsilon_{ijk} K^k, \hspace{10pt}  [J_i, K_j] = 0.\]
\eject\noindent There is one non-trivial Casimir operator, associated to the Killing form, which we may take to be:
\[ \underline{J}.\underline{J} = (t \underline{\nabla}  - \underline{x} \partial_t - \underline{x} \times \underline{\nabla}).(t \underline{\nabla}  - \underline{x} \partial_t - \underline{x} \times \underline{\nabla})\]
\[  = t^2 \underline{\nabla}.\underline{\nabla}  - (3 + \underline{x}.\underline{\nabla})t\partial_t - \underline{x}.\underline{\nabla}(1 +  t\partial_t) +\underline{x}.\underline{x} \partial_t^2 + (\underline{x} \times \underline{\nabla}).(\underline{x} \times \underline{\nabla})\]
\[  = t^2 \underline{\nabla}.\underline{\nabla}  - (3 + 2\underline{x}.\underline{\nabla})t\partial_t - \underline{x}.\underline{\nabla}  +\underline{x}.\underline{x} \partial_t^2 - ((\underline{x} \times \underline{\nabla})\times (\underline{\nabla})). \underline{x}\]
\[  = t^2 \underline{\nabla}.\underline{\nabla}  - (3 + 2\underline{x}.\underline{\nabla})t\partial_t - \underline{x}.\underline{\nabla}  +\underline{x}.\underline{x} \partial_t^2 + \underline{x}  (\underline{\nabla}.\underline{\nabla}). \underline{x} - (\underline{x}.\underline{\nabla}) (3 + \underline{x}.\underline{\nabla})\]
\[  = (t^2 + \underline{x}.\underline{x})(\partial_t^2 +  \underline{\nabla}.\underline{\nabla})  - (2 + 2\underline{x}.\underline{\nabla})t\partial_t  - (t\partial_t)^2 - (\underline{x}.\underline{\nabla}) (2 + \underline{x}.\underline{\nabla})\]
\[  = (t^2 + \underline{x}.\underline{x})(\partial_t^2 + \underline{\nabla}.\underline{\nabla}) - 2(t\partial_t + \underline{x}.\underline{\nabla}) - (t\partial_t + \underline{x}.\underline{\nabla})^2\]
\[ =  (g_{ab}x^ax^b) g^{cd} \partial_c\partial_d -  x^a \partial_a(x^b \partial_b + 2). \]
Here $x^a = (t, x, y, z)$ and $g_{ab}$ is represented by the identity matrix, so we have:
\[  g_{ab}x^ax^b = t^2 + x^2 + y^2 + z^2 = t^2 + \underline{x}.\underline{x}, \]
\[   g^{cd} \partial_c\partial_d  =  \partial_t^2 + \partial_x^2  + \partial_y^2  + \partial_z^2  = \partial_t^2 + \underline{\nabla}.\underline{\nabla}, \]
\[ x^a \partial_a = t\partial_t + x\partial_x + y\partial_y + z\partial_z = t\partial_t + \underline{x}.\underline{\nabla}.\]
Note that $\underline{K}.\underline{K}$ is the same operator as $\underline{J}.\underline{J}$.\\\\
A harmonic polynomial of degree $k$, $p(x^a)$, on $\mathbb{SU}(2, \mathbb{C})$ is the restriction to the three-sphere $g_{ab}x^a x^b = 1$ of a polynomial in $x^a$ that is harmonic $g^{ab} \partial_a \partial_b  p = 0$ and is homogeneous of non-negative integral degree $k$: $x^a \partial_a p = k p$.   So we see that a harmonic polynomial is an eigen-state of $\underline{J}.\underline{J}$  of eigen-value  $- k(k + 2)$.    The polynomial is then said to have spin $\frac{k}{2}$.   Finally by the Peter-Weyl theorem a smooth function on the group may be uniquely decomposed into its harmonic components.     Explicitly, the harmonic transformation $H(f)$ of $f(x)$ is:
\[ H(f)(y) = \int_{\mathbb{S}^3}  \frac{f(x)\omega_x}{(x - y).(x- y)}, \hspace{10pt} y.y < 1.\]
Here the dot product is the standard Euclidean dot product of vectors in $\mathbb{R}^4$.  Also the integral is taken over the unit three-sphere, $\mathbb{S}^3$, with equation $x.x = 1$ and $\omega_x$ is the invariant volume form of the three-sphere, normalized so that the integral of the constant function $1$ over the three-sphere gives the result $1$.   The homogeneous components of the Taylor expansion of $H(f)(y)$, in powers of the vector variable $y^a$, give the various spherical harmonics of $f$. Note that $\partial_y.\partial_y H(f)(y) =0$, so each component is a harmonic polynomial of the form $h^{a_1a_2\dots a_k}y_{a_1}y_{a_2}\dots y_{a_k}$ where the (constant) tensor $h^{a_1a_2\dots a_k}$ is totally symmetric and trace-free on any pair of indices. 
\section{The spinor approach to $\mathbb{SU}(2, \mathbb{C})$}   We work with a complex two-dimensional vector space $\mathbb{S}$,  called the spin space.   $\mathbb{S}$ is equipped with a quaternionic conjugation: a conjugate complex linear map from $\mathbb{S}$ to itself, whose square is the negative of the identity.  If $\alpha \in \mathbb{S}$, then its conjugate is denoted $\overline{\alpha}$.  If $\alpha \in \mathbb{S}$, then $\{ \alpha, \overline{\alpha}\}$ forms a basis for $\mathbb{S}$, over $\mathbb{C}$, if and only if $\alpha$ is non-zero.   We use upper case Latin indices for the tensor algebra over $\mathbb{C}$ of $\mathbb{S}$.   Then the conjugate of the spinor $\alpha^A$ is the spinor $\overline{\alpha}^A$.   We give $\mathbb{S}$ a complex symplectic structure, denoted $\epsilon^{AB}$ which is self-conjugate.  If $\alpha^A$ is any non-zero spinor, we have $2\alpha^{[A} \overline{\alpha}^{B]} = k \epsilon^{AB}$, where $k$ is real and non-zero.  It is easily seen that the sign of $k$ is independent of the choice of the spinor $\alpha$. We say that $\epsilon^{AB}$ is positively oriented, relative to the given conjugation, if and only if $k$ is always positive.  Note that if $\epsilon^{AB}$ is not positively oriented, then $- \epsilon^{AB}$ is positively oriented.  Henceforth we take $\epsilon^{AB}$ to be positively oriented. Then $\alpha$ is said to be normalized if and only if $k = 1$. In index-free notation, $\epsilon$ is positively oriented if and only if there exists a non-zero spinor $\alpha$, such that $2\alpha\wedge \overline{\alpha} = \epsilon$.  Then any such spinor $\alpha$ is normalized.  If $\alpha$ is normalized and $\beta$ is any spinor, then we have $\beta = p \alpha + q\overline{\alpha}$, for some complex numbers $p$ and $q$. Then $\overline{\beta} = - \overline{q} \alpha + \overline{p} \,\overline{\alpha}$ and $\beta\wedge \overline{\beta} =(|p|^2 + |q|^2) \alpha\wedge \overline{\alpha}$, so $\beta$ is itself normalized if and only if $|p|^2 + |q|^2 =1 $.   The matrix $m$ of the transformation $\alpha\rightarrow \beta$ is:
\[m = \hspace{4pt} \begin{array}{|rr|} p &-\overline{q}\\q&\overline{p}\end{array}\hspace{3pt}, \hspace{4pt} \det(m)  = |p|^2 + |q|^2 = 1.\]
The group of all such transformations is the group $\mathbb{SU}(2, \mathbb{C})$.  Using indices, the elements of $\mathbb{SU}(2, \mathbb{C})$ are represented by endomorphisms $m_A^{\hspace{4pt} B}$ which are self-conjugate and preserve $\epsilon^{AB}$: 
\[ \overline{m}_A^{\hspace{4pt} B} = m_A^{\hspace{4pt} B}, \hspace{5pt} m_C^{\hspace{4pt} A} m_D^{\hspace{4pt} B} \epsilon^{CD} = \epsilon^{AB}.\] 
Spinor indices are raised and lowered using $\epsilon^{AB}$ and its inverse $\epsilon_{AB}$, as appropriate according to the rules $v_A = v^B \epsilon_{BA}$ and $v^A = \epsilon^{AB} v_B$.   In particular, we have $\epsilon_{AB} \epsilon^{AB} =  2$.  Also a spinor $\alpha^A$ is normalized if and only if $\alpha_A\overline{\alpha}^A = 1$.
\eject\noindent
The Lie algebra of $\mathbb{SU}(2, \mathbb{C})$ is represented by self-conjugate vector fields $J_{CD} = J_{DC} $ on the group,  which obey the relations at any point $m_A^{\hspace{4pt} B}$ of the group:
\[ 2J_{CD} m_A^{\hspace{4pt} B} = - m_{A(C} \delta_{D)}^{\hspace{7pt} B}. \]
For any spinors $\alpha$ and $\beta$, put $J_\alpha = \alpha^A\alpha^B J_{AB}$, $J_\beta = \beta^A\beta^B J_{AB}$ and $J_{\alpha\beta} =  \alpha^A \beta^B J_{AB} = J_{\beta\alpha}$.   Then we have, for any spinors $\alpha$ and $\beta$:
\[ 2J_\alpha m_{A}^{\hspace{4pt} B} =  m_{A}^{\hspace{4pt} C}\alpha_C \alpha^B,\]\[   4J_\alpha J_\beta  m_{A}^{\hspace{4pt} B} =  2J_\alpha m_{A}^{\hspace{4pt} C}\beta_C \beta^B   =  - \alpha_D\beta^D m_{A}^{\hspace{4pt} C}\alpha_C\beta^B, \]
\[ 2(J_\alpha J_\beta - J_\beta J_\alpha)  m_{A}^{\hspace{4pt} B} =\alpha_D\beta^D m_{AC}(\alpha^{(C}\beta^{B)}  =  2\alpha_E\beta^E  J_{CD}\alpha^C \beta^D m_A^{\hspace{4pt}B}, \]
\[ [J_\alpha,  J_\beta] =  \alpha_E\beta^E  J_{\alpha \beta}.\]
Here the bracket denotes the Lie bracket of vector fields on the group.
Polarizing appropriately, with respect to $\alpha$ and $\beta$, we get also:
\[ 2[J_\alpha,  J_{\alpha\beta}] =  \alpha_E\beta^E  J_{\alpha}, \hspace{10pt}  2[J_\beta,  J_{\alpha\beta}] =   -  \alpha_E\beta^E  J_{\beta}. \]
Take $\alpha$ to be normalized and put $\beta = \overline{\alpha}$, $J_3 = iJ_{\alpha\overline{\alpha}}$, $J_+ = J_\alpha$ and $J_- = J_{\overline{\alpha}}$.   Then we have:
\[ [J_{+}, J_-]  = J_3, \hspace{10pt} 2[J_\pm, J_3] = \pm  J_\pm,\]
\[ J_{CD} = \alpha_C \alpha_D J_- + \overline{\alpha}_C \overline{\alpha}_D J_+ + 2i \alpha_{(C} \overline{\alpha}_{D)} J_3.\] 
Finally write $J_\pm = J_1 \pm i J_2$ and put $\underline{J} = [J_1, J_2, J_3]$.  Then $J_1$, $J_2$ and $J_3$ are real vector fields and we have:
\[  2[J_i, J_j] = \epsilon_{ijk} J_k.\]
Then we have:  $J_{CD}J^{CD}  =  J_- J_+ + J_+ J_-  + 2J_3^2 = 2(J_1^2 + J_2^2 + J_3^2) = 2\underline{J}.\underline{J}$.
Comparing with our previous $\underline{J}$ operator, which obeyed the commutation relations $[J_i, J_j] = 2\epsilon_{ijk} J_k$, we see that if we were to multiply each of our present operators by a factor of four, the commutation relations would be the same.   So the operator $J_{CD} J^{CD}$ is one-eighth of our previous Casimir operator, whose eigen-values were $- k(k + 2)$ with $k$ a non-negative integer.  Thus the operator $J_{CD}J^{CD}$ has eigen-values $- \frac{1}{8} k(k + 2)$.   In particular, acting on $m_A^{\hspace{4pt}B}$ we have:
\[ 8J_{CD}J^ {CD} m_A^{\hspace{4pt} B} = - 4J_{CD}m_{A}^{\hspace{4pt} C} \epsilon^{DB} = 2 m_{A(C} \delta_{D)}^{\hspace{7pt} C} \epsilon^{DB} = - 3 m_A^{\hspace{4pt} B}.\] 
So $k = 1$ and we say that $m_A^{\hspace{4pt} B}$ is (pure) spin one-half (the parameter $k$ corresponding to twice the spin).  Similarly, for $n$ any positive integer and  for any constant spinors $p_B$ and $q^A$,
the spinor function $(m_A^{\hspace{4pt} B}p_Bq^A)^n$ is of spin $\displaystyle{\frac{n}{2}}$. \eject\noindent
\section{The $\Xi$-transform for compact Lie groups}
Let $f(g, h)$ denote a smooth function of variables $g$ and $h$ in $\mathbb{G}$.  So $f: \mathbb{G} \times \mathbb{G} \rightarrow \mathbb{R}$.  Denote by $\nabla^+_\alpha f$ and $\Delta^+_\alpha f$ the action of the invariant vector fields $\nabla_\alpha$ and $\Delta_\alpha$ of $\mathbb{G}$ on the first argument of $f$,  respectively and by $\nabla^-_\alpha f$ and $\Delta^-_\alpha f$ the action of  the invariant vector fields on the second argument of $f$,  respectively.   If $C$ is a Casimir operator of $\mathbb{G}$, denote by $C^+$ its action on the first argument of $f(g, h)$ and  by $C^-$ its action on the second argument of $f(g, h)$.  If $C$ is of degree $k$, define its extension to two variables, denoted $\mathcal{C}$,  by the formula: \[ \mathcal{C} = C^+ - (-1)^k C^-.\]
For each smooth $f(g, h)$, define $\Xi(f)$, called the $\Xi$-transform of $f$, by the formula:
\[ \Xi(f)(g, h) = \int f(p, g^{-1}ph) \omega_p.\]
Here $\omega_p$ is the Haar measure in the $p$-variable, represented as a smooth real volume form on $\mathbb{G}$.
In these and the subsequent integrals, the variable $p \in \mathbb{G}$ is understood to range over the whole Lie group $\mathbb{G}$.
Note that using the bi-invariance of the form $\omega_p$, we can rewrite this transform in the various ways:
\[ \Xi(f)(g, h) =  \int f(p, g^{-1}ph) \omega_p =  \int f(gp, ph) \omega_p =  \int f(gph^{-1}, p) \omega_p =   \int f(ph^{-1}, g^{-1}p) \omega_p.  \]
It is clear that $\Xi(f)$ is a smooth function on $\mathbb{G} \times \mathbb{G}$.  We calculate the operators $C^\pm $ acting on $\Xi(f)(g, h)$: 
\begin{itemize} \item First multiply $h$ on the right by $\exp_v(t)$.   Then we have, for any real $t$ and any $v^\alpha$:
\[ \Xi(f)(g, h\exp_v(t)) =  \int f(p, g^{-1}ph\exp_v(t)) \omega_p\]
\[ = \int f(p, uR_v(t))_{u = g^{-1}ph} \omega_p\]
Differentiating with respect to $t$ and putting $t = 0$, we have:
\[ (\nabla^-_\alpha  \Xi(f))(g, h) = \int (\nabla^-_\alpha f)(p, u)_{u = g^{-1}ph} \omega_p.\]  
Then we have, by repeated differentiation of the last formula:
\[ (C^-\Xi(f))(g, h) = \int (C^- f)(p, u)_{u = g^{-1}ph} \omega_p.\]
\item Second, we multiply $g$ on the right by $\exp_v(t)$ in the formula for $\Xi(f)(g, h)$. Then we have, for any real $t$ and any $v^\alpha$:
\[ \Xi(f)(g\exp_v(t), h) =  \int f(p, \exp_v(-t)g^{-1}ph) \omega_p  = \int f(p, L_v(-t)u)_{u = g^{-1}ph} \omega_p\]
Differentiating with respect to $t$ and putting $t = 0$, we have:
\[ (\nabla^+_\alpha  \Xi(f))(g, h) =  - \int (\Delta^-_\alpha f)(p, u)_{u = g^{-1}ph} \omega_p.\]
Then we have, for $C$ a Casimir operator of order $k$, after iteration of this formula:
\[ (C^+ \Xi(f))(g, h) =(-1)^k \int (C^-f)(p, u)_{u = g^{-1}ph} \omega_p.\]
\end{itemize} 
Subtracting these two relations, we have proved:
\begin{itemize} \item If $C$ is a Casimir operator of $\mathbb{G}$ of order $k$, then the function $\Xi(f)(g, h)$ obeys the invariant partial differential equation of order $k$:
\[(\mathcal{C} \Xi(f))(g, h) = 0.\]
\end{itemize} 
Next we rewrite the $\Xi$-transform with the arguments $g$ and $h$ moved to the first argument of $f$:
\[  \Xi(f)(g, h)  =  \int f(gph^{-1}, p) \omega_p .\]
Then we have, as above:
\[ \Xi(f)(g, \exp_v(t)h)  =  \int f(gph^{-1}\exp_v(-t), p) \omega_p, \]
\[ (\Delta^-_\alpha \Xi(f))(g, h) =     - \int  (\nabla^+_\alpha f)(gph^{-1},  p) \omega_p, \]
\[  ( C^-\Xi(f))(g, h) =    (-1)^k \int  (C^+ f)(gph^{-1}g,  p) \omega_p =  (-1)^k \int  (C^+ f)(p,  g^{-1}ph) \omega_p.\]
But earlier we proved the relation:  $ (C^-  \Xi(f))(g, h) =  \int (C^-f)(p, g^{-1}ph)\omega_p$.
Hence, by subtraction, we have the formula:
\[ 0 = \int (C^+ - (-1)^k C^-)(f)(p, g^{-1} ph) \omega_p.\]
We have shown the key result:
\begin{itemize}  \item  For any Casimir operator $C$ on $\mathbb{G}$, the $\Xi$-transform obeys the relations: 
\[ \mathcal{C}\circ \Xi = \Xi \circ \mathcal{C} = 0.\]
\end{itemize}
\eject\noindent
\section{The kernel of the $\Xi$-transform in the $\mathbb{SU}(2, \mathbb{C}) $ case}
In this section, we make the simplifying assumption that all functions involved are linear combinations of only finitely many spherical harmonics.  We expect the results to go through without this assumption, but we do not prove this here.\\\\
Henceforth we assume that $\mathbb{G} = \mathbb{SU}(2, \mathbb{C}) $, which is also the Lie group of unit quaternions under multiplication.  Denote by $C$ its second-order Casimir operator, determined by the negative of the inverse of its Killing form, so that its eigen-values are $n(n + 2)$, for $n$ a non-negative integer.   For non-negative integers $k$ and $l$, denote by $\mathbb{H}_{k, l}$ the space of all (necessarily real analytic) harmonics $f(g, h)$ on $\mathbb{SU}(2, \mathbb{C}) \times \mathbb{SU}(2, \mathbb{C})$ that obey the relations $C^+ f = k(k + 2)f$ and $C^- f = l(l + 2) f$.   Also for $N$ any non-negative integer, put:
\[ \mathbb{H}^+_N = \oplus_{k = 0}^N \mathbb{H}_{k, k}, \hspace{10pt} \mathbb{H}^-_N = \oplus_{k = 0, l = 0, k \ne l}^N \mathbb{H}_{k, l}, \]
\[ \mathbb{H}_N = \mathbb{H}^+_N  \oplus \mathbb{H}^-_N  =  \oplus_{k = 0, l = 0}^N \mathbb{H}_{k, l}. \]
The spaces $\mathbb{H}^+_N$ and $\mathbb{H}^-_N$ are mutually orthogonal with respect to Haar measure.\\\\  Henceforth we fix $N$ and assume $f \in \mathbb{H}_N$.   Suppose that the $\Xi$-transform of $f$ vanishes, so we have $\Xi(f)(g, h)  = \int f(p, g^{-1}ph) \omega_p = 0$.  The harmonic parts $f_{k, l}$ of $f$ with $k \ne l$ are killed by the $\Xi$-transform, since such parts lie in the image of $\mathcal{C} = C^+ - C^-$, since $\mathcal{C} f_{k, l} = (k - l)(k + l + 2)f_{k, l}$, or equivalently, since the eigen-value function $x(x + 2)$ of $C$ is one-to-one for $x$ real and non-negative.   So the $\Xi$-transform kills the space $\mathbb{H}^-_N$.   Our aim is to show that nothing else lies in the kernel of $\Xi$ acting on $\mathbb{H}_N$.  Accordingly, we need only look at functions $f(g, h)$ of the form: $f = \sum_k f_k(g, h)$ where $f_k(g, h)$ obeys $C^+ f_k = C^- f_k =  k(k + 2) f_k $, so $f \in \mathbb{H}^+_N$.
Then we may write the condition that $f$ be in the kernel:  
\[ \sum_k   \int   f_k(x^a, x^b m_b^a) \omega_x = 0.\]
Here $x^a = (t, x, y, z) \in \mathbb{R}^4$ and each $f_k(x^a y^b)$ is bi-harmonic, so may be written: 
\[ f_k (x^a, y^b) = f_{a_1a_2\dots a_k}^{b_1b_2\dots b_k} x^{a_1} x^{a_2} \dots x^{a_k} y_{b_1} y_{b_2} \dots y_{b_k}.\] Here $f_{a_1a_2\dots a_k}^{b_1b_2\dots b_k}$ is  a constant tensor and is symmetric and trace-free on any pair of its upper indices and symmetric and trace-free on any pair of its lower indices.  Also the matrix $m_b^a $ is an arbitrary $\mathbb{SO}(4)$ transformation and $\omega_x$ is the invariant volume measure on the unit three-sphere: $g_{ab} x^a x^b = 1$.    Here $g_{ab}$ is the Euclidean metric tensor for $\mathbb{R}^4$.
\eject\noindent 
We can do the integral explicitly, leaving the following sum to be analyzed:
\[ 0 = \sum c_k f^{b_1b_2\dots b_k}_{a_1 a_2 \dots a_k} m^{a_1}_{b_1} m^{a_2}_{b_2} \dots m^{a_k}_{b_k}.\]
Here $c_k$ is a positive (computable) constant, depending only on $k$.  Since $m_b^a$ is a $\mathbb{SO}(4, \mathbb{R})$ transformation, it obeys the equation:
\[ m^{\hspace{4pt}a}_b m^{\hspace{4pt}c}_d g_{ac} = g_{bd}.\]
We can differentiate this relation with vector fields $ J_{ca} = m_{e[c}\partial_{a]}^e$.  We have:
\[ m_{ec}\partial_a^e m^{\hspace{4pt}p}_b m^{\hspace{4pt}q}_d g_{pq}  =  m_{ec}\delta_a^p \delta^e_b m^{\hspace{4pt}q}_d g_{pq} + m_{ec}\delta_a^q \delta^e_d m^{\hspace{4pt}p}_b  g_{pq}   =  m_{bc}  m_{da}  + m_{dc}  m_{ba}\]
So taking the skew part in the indices $ca$, we find that $J_{ca}$ kills the defining relation for $\mathbb{SO}(4, \mathbb{R})$.  Then $J_{ca}$ gives the Lie algebra of $\mathbb{SO}(4, \mathbb{R})$.  Also we have:
\[ J_{cd} m_a^{\hspace{4pt} b} = m_{e[c}\delta_{d]}^{\hspace{6pt} b} \delta^e_a = m_{a[c}\delta_{d]}^{\hspace{6pt} b} .\]
Now we differentiate our sum, using the Lie algebra operators $J_{cd}$.  The result is:
\[ 0 = \sum  k c_k m_{b_1[c} f^{b_1b_2\dots b_k}_{d] a_2 \dots a_k} m^{a_2}_{b_2} \dots m^{a_k}_{b_k}.\]
We differentiate again, using the generators $J_{ef}$, giving the formula:
\[ 0 = \sum  k(k - 1) c_k m_{b_2[e}m_{|b_1|[c} f^{b_1b_2b_3\dots b_k}_{d] f] a_3\dots a_k} \dots m^{a_k}_{b_k} +   kc_k (J_{ef} m_{b_1[c}) f^{b_1b_2\dots b_k}_{d] a_2 \dots a_k} m^{a_2}_{b_2} \dots m^{a_k}_{b_k}\] 
\[ = \sum  k(k - 1) c_k m_{b_2[e}m_{|b_1|[c} f^{b_1b_2b_3\dots b_k}_{d] f] a_3\dots a_k} m^{a_3}_{b_3} \dots m^{a_k}_{b_k} +   kc_k m_{b_1[e} g_{f][c} f^{b_1b_2\dots b_k}_{d] a_2 \dots a_k} m^{a_2}_{b_2} \dots m^{a_k}_{b_k}.\] 
Trace this relation with $- 2g^{ec} g^{fd}$:
\[   0 = \sum  -2 g^{ec}g^{fd} k(k - 1) c_k m_{b_2e}m_{|b_1|[c} f^{b_1b_2b_3\dots b_k}_{d] f a_3\dots a_k} m^{a_3}_{b_3} \dots m^{a_k}_{b_k} -2 g^{ec}g^{fd}   kc_k m_{b_1e} g_{f[c} f^{b_1b_2\dots b_k}_{d] a_2 \dots a_k} m^{a_2}_{b_2} \dots m^{a_k}_{b_k}\] 
\[   = \sum  - 2g^{fd} k(k - 1) c_k m_{b_2}^{\hspace{6pt}c}m_{|b_1|[c} f^{b_1b_2b_3\dots b_k}_{d] f a_3\dots a_k} m^{a_3}_{b_3} \dots m^{a_k}_{b_k} -  2kc_k m_{b_1}^{\hspace{6pt}c} \delta_{[c}^d f^{b_1b_2\dots b_k}_{d] a_2 \dots a_k} m^{a_2}_{b_2} \dots m^{a_k}_{b_k}\] 
\[   =  \sum  k(k + 2)c_k  f^{b_1b_2b_3\dots b_k}_{a_1 a_2 a_3\dots a_k} m_{b_1}^{\hspace{6pt}a_1} m_{b_2}^{\hspace{6pt}a_2}m^{a_3}_{b_3} \dots m^{a_k}_{b_k} .\]
So each term $c_k f^{b_1b_2\dots b_k}_{a_1 a_2 \dots a_k} m^{a_1}_{b_1} m^{a_2}_{b_2} \dots m^{a_k}_{b_k} $ is an eigenstate of the second order $\mathbb{SO}(4)$-Casimir operator $-2g^{ec} g^{fd} J_{cd} J_{ef} = - 2J_{cd}J^{cd}$ of eigenvalue $k(k + 2)$.  Eigenstates of different eigen-values are orthogonal with respect to the Haar measure of $\mathbb{SO}(4)$.  Since the function $k(k + 2)$ is single-valued, when $k$ is non-negative, the various terms in the sum are mutually orthogonal.  Since the sum is zero, each individual term must be zero, giving the relation, for each non-negative integer $k$:   
\[ 0 =  f^{b_1b_2\dots b_k}_{a_1 a_2 \dots a_k} m^{a_1}_{b_1} m^{a_2}_{b_2} \dots m^{a_k}_{b_k}.\]
\eject\noindent
We would like to conclude that $ f^{b_1b_2\dots b_k}_{a_1 a_2 \dots a_k} $ vanishes identically.   To do this we may henceforth assume that $k \ge 1$.  We re-write the relation using spinors and the isomorphism of the spin group of $\mathbb{SO}(4)$ (a double cover of $\mathbb{SO}(4)$) with $\mathbb{SU}(2, \mathbb{C})\times \mathbb{SU}(2, \mathbb{C})$:
\[ 0 = f^{B_1B_2\dots B_kB_1'B_2'\dots B_k'}_{A_1 A_2 \dots A_kA_1'A_2'\dots A_k'} m^{\hspace{4pt} A_1}_{B_1} m^{\hspace{4pt} A_2}_{B_2} \dots m^{\hspace{4pt} A_k}_{B_k} m^{\hspace{4pt} A_1'}_{B_1'} m^{\hspace{4pt} A_2'}_{B_2'} \dots m^{\hspace{4pt} A_k'}_{B_k'}.\]
Here $m_A^{\hspace{4pt} B}$ and $m_{A'}^{\hspace{4pt} B'}$ are elements of $\mathbb{SU}(2, \mathbb{C})$, independent of each other.   So it is sufficient to prove:
\[ 0 = f^{B_1B_2\dots B_k}_{A_1 A_2 \dots A_k} m^{\hspace{4pt} A_1}_{ B_1} m^{\hspace{4pt} A_2}_{B_2} \dots m^{\hspace{4pt} A_k}_{B_k} \implies f^{B_1B_2\dots B_k}_{A_1 A_2 \dots A_k} = 0.\]
Here we may assume, without loss of generality,  that $f^{B_1B_2\dots B_k}_{A_1 A_2 \dots A_k} $ is symmetric under the simultaneous interchange of an $(A_p, B_p)$-pair with an $(A_q, B_q)$-pair, for any $p$ and $q$.  Then $ f^{B_1B_2\dots B_k}_{A_1 A_2 \dots A_k} $ has $\binom{k+3}{3} = \frac{1}{6}(k + 3)(k + 2)(k + 1)$ independent components.
Decomposing this relation into irreducible representations in the upper indices, using the fact that $ m_{[A}^{\hspace{6pt}C}m_{B]}^{\hspace{6pt}D} = \delta_{[A}^{\hspace{6pt}C}\delta_{B]}^{\hspace{6pt}D}$, since $\mathbb{SU}(2, \mathbb{C})$ transformations have unit determinant, the relation becomes:
\[ 0 = \sum_{r =0}^{k}  g^{B_1B_2\dots B_r}_{A_1 A_2 \dots A_r} m^{\hspace{4pt} A_1}_{B_1} m^{\hspace{4pt} A_2}_{B_2} \dots m^{\hspace{4pt} A_r}_{B_r}.\]
Here $g^{B_1B_2\dots B_r}_{A_1 A_2 \dots A_r} $ is obtained from $f^{B_1B_2\dots B_k}_{A_1 A_2 \dots A_k} $ by tracing over pairs $(A_1, A_2)$, $(B_1, B_2)$ \dots $(A_{2s-1}, A_{2s})$, $(B_{2s-1}, B_{2s})$ with a skew spinor symplectic form $\epsilon_{CD}$, or its inverse $\epsilon^{CD}$, as appropriate (here $r + 2s = k$, so $k - r$ is necessarily even), symmetrizing over the remaining $B$-indices and over the remaining $A$-indices and multiplying by a suitable positive constant, depending only on $r$.  We have $(r + 1)^2$ independent components for each $g^{B_1B_2\dots B_r}_{A_1 A_2 \dots A_r}$, in agreement with the simple combinatorial identities:
\begin{itemize} \item When $k = 2s$ is even:
\[ \binom{2s+3}{3} = \frac{1}{6} (2s + 3)(2s + 2)(2s + 1) =  1^2 + 3^2 + 5^2 + \dots (2s + 1)^2.\]
\item When $k = 2s + 1$ is odd. 
\[  \binom{2s+4}{3} = \frac{1}{6} (2s + 4)(2s + 3)(2s + 2) =  2^2 + 4^2 + \dots + (2s)^2 + (2s + 2)^2.\]
\end{itemize}
We now need to show that necessarily each  $g^{B_1B_2\dots B_r}_{A_1 A_2 \dots A_r} $ vanishes.    We may further decompose into irreducible spinor representations, so the required sum now takes the form:
\[ 0 = \sum_{0 \le p, q \le k} (h_{p, q})^{B_1B_2\dots B_p}_{A_1 A_2 \dots A_p}m^{\hspace{4pt} A_1}_{B_1} m^{\hspace{4pt} A_2}_{B_2} \dots m^{\hspace{4pt} A_p}_{B_p} (m^{\hspace{4pt} C}_{ C})^q.\]
Here the quantities $(h_{p, q})^{B_1B_2\dots B_p}_{A_1 A_2 \dots A_p}$ for $p + q = r$ are the irreducible spinor parts of the spinor $g^{B_1B_2\dots B_r}_{A_1 A_2 \dots A_r}$, so $(h_{p, q})^{B_1B_2\dots B_p}_{A_1 A_2 \dots A_p}$  is trace-free on any index pair $(A_j, B_k)$ and is symmetric in its upper indices and symmetric in its lower indices. We need only to prove that each $(h_{p, q})^{B_1B_2\dots B_p}_{A_1 A_2 \dots A_p}$ vanishes identically.  \\\\
Note that the matrix $m_A^B$ may be written uniquely as follows:
\[ m_A^{\hspace{4pt} B} = \hspace{4pt} \begin{array}{|cc|} t + iz & x + iy\\- x + iy& t - i z\end{array}\hspace{3pt} .\]
Here we have $(t, x, y, z) \in \mathbb{R}^4$ and $t^2 + x^2 + y^2 + z^2 = t^2 + \underline{x}.\underline{x} = 1$.  Also $\underline{x} = (x, y, z)$ lies in $\mathbb{R}^3$, equipped with its usual Euclidean dot product.
Note that we have also  $m_A^{\hspace{4pt} A} = 2t = \pm 2\sqrt{1 - \underline{x}.\underline{x}}$.   In this language, we now need to analyze the relation, valid whenever $\underline{x}.\underline{x} \le 1$:
\[ 0 = \sum_{p, q} 2^q t^q h_{p, q}( \underline{x}).\]
Here $\underline{x} \in \mathbb{R}^3$ and the function $h_{p, q}( \underline{x})$ is a polynomial: $(h_{p, q})_{a_1a_2 \dots a_p} x^{a_1} x^{a_2} \dots x^{a_p}$, where $1\le a_j \le 3$, for each $j$.  The coefficient tensor $(h_{p, q})_{a_1a_2 \dots a_p}$ is totally symmetric and trace-free on any index pair.  First put $\underline{x}.\underline{x} = 1$, which entails that $t = 0$, giving the formula:
\[ 0 = \sum_{p}  h_{p, 0}( \underline{x}), \hspace{10pt} \underline{x}.\underline{x} = 1.\]
But, as $p$ varies, the functions $h_{p, 0}( \underline{x})$ are spherical harmonics on the two-sphere  $\underline{x}.\underline{x} = 1$, of different eigenvalues with respect to the Laplacian operator of the two-sphere, so are mutually orthogonal.  So $h_{p, 0}( \underline{x}) = 0$ on the two-sphere, for each $p$.  But $h_{p, 0}(\underline{x})$ is homogeneous in $\underline{x}$ of degree $p$, so $h_{p, 0}(x)$ vanishes identically on $\mathbb{R}^3$, so $(h_{p, 0})^{B_1B_2\dots B_p}_{A_1 A_2 \dots A_p}$ vanishes identically, for each $p$.  Then the remaining sum has a factor of $2t$, so we may factor out and reduce to the relation:
\[ 0 = \sum_{p, q} 2^q t^q h_{p, q+1}(\underline{x}).\]
Again put $\underline{x}.\underline{x} = 1$, so $t =0$, giving the formula:
\[ 0 = \sum_{p}  h_{p, 1}( \underline{x}), \hspace{10pt} \underline{x}.\underline{x} = 1.\]
As before we conclude that each $(h_{p, 1})^{B_1B_2\dots B_p}_{A_1 A_2 \dots A_p}$ vanishes identically.  Now we iterate and conclude that each $(h_{p, q})^{B_1B_2\dots B_p}_{A_1 A_2 \dots A_p}$ vanishes identically, as required, for all $p$ and $q$ and we are done.
Summarizing, we have given a proof of the following result:
\begin{itemize}\item The kernel of $\Xi$ acting on the space $\mathbb{H}_N$ is the image of $\mathcal{C}$ acting on the same space. 
\end{itemize}
Finally, we notice that in the course of this proof, we have found that the image of $\Xi$ acting on $\mathbb{H}_N$ does not involve harmonics apart from those of the form $f_{k,k}$ for $0 \le k \le N$.   This follows from the relation $C^+ = C^- = - 2J_{cd}J^{cd}$, acting on functions of the form:  $f^{b_1b_2\dots b_k}_{a_1 a_2 \dots a_k} m^{a_1}_{b_1} m^{a_2}_{b_2} \dots m^{a_k}_{b_k} $.  We describe this first, before finishing our argument. 
\\\\
We compare our normalizations of the Casimir operators for the groups $\mathbb{SO}(4,\mathbb{R} )$ and $\mathbb{SU}(2,\mathbb{ C})\times \mathbb{SU}(2, \mathbb{C})$ as follows.   Using spinors the group elements $m_a^{\hspace{4pt}b}$  of $\mathbb{SO}(4,\mathbb{R} )$ decompose as products of $\mathbb{SU}(2,\mathbb{ C})$ elements:
\[ m_a^{\hspace{4pt}b} = m_A^{\hspace{4pt} B} m_{A'}^{\hspace{4pt}B'}.\]
The Lie algebra operator $J_{cd} = - J_{dc}$ decomposes as:
\[ J_{cd} = \epsilon_{CD} J_{C'D'} + \epsilon_{C'D'} J_{CD}.\]
Here $J_{CD} = J_{DC}$ and $J_{C'D'}  = J_{D'C'}$ generate the independent factors of  $\mathbb{SU}(2, \mathbb{C})$. 
Acting on $m_a^{\hspace{4pt}b}$, we get:
\[ 2J_{CD} m_a^{\hspace{4pt}b} =  \epsilon^{C'D'} J_{cd} m_a^{\hspace{4pt}b} =  \epsilon^{C'D'}  m_{a[c} \delta_{d]}^{\hspace{7pt}b} =  \epsilon^{C'D'}  m_{A'C'} m_{A(C} \delta_{D)}^{\hspace{7pt}B} \delta_{D'}^{\hspace{4pt}B'} =  -   m_{A'}^{\hspace{4pt}B'} m_{A(C} \delta_{D)}^{\hspace{7pt}B}. \]
So we have:
\[ J_{CD} m_{A'}^{\hspace{4pt} B'} = 0, \hspace{10pt} J_{CD} m_{A}^{\hspace{4pt} B} =   - \frac{1}{2} m_{A(C} \delta_{D)}^{\hspace{7pt}B}, \]
\[ J_{C'D'} m_{A}^{\hspace{4pt} B} = 0, \hspace{10pt} J_{C'D'} m_{A'}^{\hspace{4pt} B'} =   - \frac{1}{2} m_{A'(C'} \delta_{D')}^{\hspace{7pt}B'}. \]
Now we can compare Casimir operators.   We have:
\[ J^{cd} J_{cd} m_{ab} = J^{cd}   m_{a[c}g_{d]b} =  J_{cb} m_a^{\hspace{4pt} c} =  m_{a[c} \delta_{b]}^{\hspace{7pt}c}  = - \frac{3}{2} m_{ab}, \] 
\[ J^{CD} J_{CD} m_{A}^{\hspace{4pt} B} =   - \frac{1}{2} J^{CD} m_{A(C} \delta_{D)}^{\hspace{7pt}B}  = \frac{1}{2}\epsilon^{BD} J_{CD} m_{A}^{\hspace{4pt}C} = -\frac{1}{4}\epsilon^{BD}   m_{A(C} \delta_{D)}^{\hspace{7pt}C} = - \frac{3}{8} m_A^{\hspace{4pt} B}. \]
Similarly we have $\displaystyle{J^{C'D'} J_{C'D'} m_{A'}^{\hspace{4pt} B'} = - \frac{3}{8}  m_{A'}^{\hspace{4pt} B'}} $. 
The group elements $m_A^{\hspace{4pt}B}$ are of spin one-half, so the second-order Casimir operator $C$, acting on $m_A^{\hspace{4pt} B}$,  has the eigen-value $1(1 + 2) = 3$. So we have $J_{CD} J^{CD} = - \frac{1}{8} C^+$.  Similarly, we have also $J_{C'D'} J^{C'D'} = - \frac{1}{8} C^-$. These equations are consistent with the identity $J_{cd}J^{cd} =  2(J_{CD}J^{CD} + J_{C'D'} J^{C'D'})$, which follows immediately from the definition of the spinor operators $J_{CD}$ and $J_{C'D'}$ in terms of $J_{cd}$.  Acting on polynomials in $m_a^{\hspace{4pt} b}$ we then have:
\[ J_{CD}J^{CD} = J_{C'D'} J^{C'D'} = \frac{1}{4} J_{cd}J^{cd}.\]
So, finally we have $C^+ = C^- = - 2J_{cd}J^{cd}$.   In particular the various polynomials $f^{b_1b_2\dots b_k}_{a_1 a_2 \dots a_k} m^{a_1}_{b_1} m^{a_2}_{b_2} \dots m^{a_k}_{b_k} $ used above, which we showed are eigen-states of $- 2J_{cd}J^{cd}$ of eigen-value $k(k + 2)$ are also eigen-states of $C^+$ and $C^-$, each with the same eigen-value $k(k + 2)$.  \\\\
So acting on $\mathbb{H}_N$, the $\Xi$-transform annihilates $\mathbb{H}^-_N$ and has zero kernel acting on $\mathbb{H}_N^+$.   But we have just shown that the image of each $\mathbb{H}_{k, k}$ under $\Xi$ lies in $\mathbb{H}_{k, k}$.  Since the kernel is zero and each $\mathbb{H}_{k, k}$ is finite dimensional, the restriction of $\Xi$ to each $\mathbb{H}_{k, k}$ is an isomorphism with its image.  So the kernel of $\Xi$ acting on $\mathbb{H}_N$ is precisely the space $\mathbb{H}^-_N$ and the image of $\Xi$ acting on $\mathbb{H}_N$ is precisely the space $\mathbb{H}^+_N$.   The latter space is also the kernel of the operator $\mathcal{C} = C^+ - C^-$ acting on $\mathbb{H}_N$.   We have proved, for any non-negative integer $N$: 
\begin{itemize}
\item The kernel of $\mathcal{C}$ acting on the space $\mathbb{H}_N$  is the image of $\Xi$, acting on the same space.  
\item The kernel of $\Xi$ acting on the space $\mathbb{H}_N$  is the image of $\mathcal{C}$, acting on the same space.  
\end{itemize}

\eject\noindent
\section{The null twistor geometry}
The twistor space for conformally flat space-time may be taken to be a four dimensional complex vector space $\mathbb{T}$, whose elements are called twistors \cite{pen5}-\cite{pen1}.  Complexified, conformally compactified space-time is recovered as the Grassmanian $\textrm{Gr}(2, \mathbb{T})$ of all two-dimensional complex subspaces of $\mathbb{T }$, with the conformal structure such that  $x$ and $y$ in $\textrm{Gr}(2, \mathbb{T})$ are null related if and only if they have a non-zero twistor in common.   The space $\mathbb{T}$ is equipped with a pseudo-hermitian structure of signature $(2, 2)$.    If $Z^\alpha $ denotes a twistor, then its conjugate is denoted $\overline{Z}_\alpha$, which lies in the complex dual space $\mathbb{T}^*$ of $\mathbb{T}$. Then the (real ) inner product of $Z^\alpha$ with itself is $Z^\alpha \overline{Z}_\alpha$.   The space $\mathbb{T}$ is the disjoint union of three sets, $\mathbb{T}^\pm = \{ Z^\alpha \in \mathbb{T}: \pm Z^\alpha \overline{Z}_\alpha > 0\}$ and $\mathbb{N} = \{ Z^\alpha \in \mathbb{T}: Z^\alpha \overline{Z}_\alpha = 0\}$.   The twistors of $\mathbb{N}$ are called null.  Put $\mathbb{N}' = \mathbb{N} - \{0\}$.  Then $\mathbb{N}'$ is a smooth real manifold of dimension seven.   We say that $x \in  \textrm{Gr}(2, \mathbb{T})$ is real if and only if  $x \subset \mathbb{N}$. The key fact relating twistor theory to space-time is that the subset $\mathbb{X}$ of all real elements of $\textrm{Gr}(2, \mathbb{T})$ is a smooth four-manifold, equipped with the induced natural structure conformal structure, which makes $\mathbb{X}$ a conformal compactification of Minkowksi space-time.  Each element $Z$ in $\mathbb{N}'$ belongs to a one-parameter family $\gamma(Z)$ (a circle) of elements of $\mathbb{X}$, which forms a null geodesic in $\mathbb{X}$ and every null geodesic arises in this way.   Also $\gamma(Z) = \gamma(Z')$  if and only if $Z' = \lambda Z$, for $0 \ne \lambda\in \mathbb{C}$.    \\\\The twistor space $\mathbb{T}$ carries a natural flat pseudo-K\"ahler metric $g = dZ^\alpha d\overline{Z}_\alpha$, whose signature is $(4, 4)$.   Restricting to the space $\mathbb{N}' $, this metric degenerates, with the direction of degeneracy given by the homogeneity vector field $H = Z^\alpha \partial_\alpha + \overline{Z}_\alpha \overline{\partial}^\alpha$, where $\displaystyle{\partial_\alpha = \frac{\partial}{\partial Z^\alpha}}$ and $ \displaystyle{\overline{\partial}^\alpha = \frac{\partial}{\partial \overline{Z}_\alpha}}$.  The one parameter group for this vector field is the transformation $Z^\alpha \rightarrow e^tZ^\alpha$ with $t$ real.   Quotienting  $\mathbb{N}' $ out by this vector field, we get a smooth six-manifold, denoted $\mathbb{M}$, with  a conformally flat conformal structure of signature $(3, 3)$, naturally induced by the pseudo-K\"ahler structure of $\mathbb{T}$.    We may assign co-ordinates $Z = (\alpha, \beta, \gamma, \delta) \in \mathbb{C}^4$ for $\mathbb{T}$, such that:
\[ Z^\alpha \overline{Z}_\alpha = |\alpha|^2 + |\beta|^2 - |\gamma|^2 - |\delta|^2.\]
Then the twistors of $\mathbb{N}'$ satisfy $ |\alpha|^2 + |\beta|^2 = |\gamma|^2 + |\delta|^2 > 0$.  When we quotient by the scaling $Z \rightarrow e^t Z$, we may take $ |\alpha|^2 + |\beta|^2 = |\gamma|^2 + |\delta|^2 = 1$, so $\mathbb{M}$ is the product $\mathbb{S}^3\times \mathbb{S}^3$, where $\mathbb{S}^3$ is the real three-sphere.    Further the induced metric is the metric $ |d\alpha|^2 + |d\beta|^2 - |d\gamma|^2 - |d\delta|^2$ which gives $\mathbb{M}$ the metric $G_+ - G_-$ (of signature $(3, 3)$), where the $G_\pm$ are unit three-sphere metrics, applied to the first  and second factors of the product  $\mathbb{S}^3\times \mathbb{S}^3$ in the cases of $G_+$ and $G_-$, respectively.\eject\noindent
\section{The twistor symmetry Lie groups and algebras }
The symmetry group $\mathbb{O}(\mathbb{T})$ of $\mathbb{T}$, equipped with its pseudo-K\"ahler structure, is the group of real linear transformations of $\mathbb{T}$ to itself preserving the pseudo-K\"ahler metric.  Then $\mathbb{O}(\mathbb{T})$ is isomorphic to the twenty-eight dimensional real Lie group $\mathbb{O}(4, 4, \mathbb{R})$.   The subgroup $\mathbb{U}(\mathbb{T})$ of $\mathbb{O}(\mathbb{T})$ is the subgroup that preserves the complex structure.  It has sixteen real dimensions and is isomorphic to the Lie group $\mathbb{U}(2, 2, \mathbb{C})$.  The Lie algebra $\frak{o}(\mathbb{T})$ of $\mathbb{O}(\mathbb{T})$  is represented on the twistor space by the operators: 
\[ E^{\alpha\beta} = 2Z^{[\alpha} \overline{\partial}^{\beta]}, \hspace{10pt}  \overline{E}_{\alpha\beta}  = 2\overline{Z}_{[\alpha} \partial_{\beta]}, \hspace{10pt} E_\beta^\alpha = Z^\alpha \partial_\beta - \overline{Z}_\beta \overline{\partial}^\alpha.\]
The operators $E^{\alpha\beta}$ and $\overline{E}_{\alpha\beta}$ have six complex degrees of freedom, counting for twelve real dimensions.   The operator $E^\alpha_\beta$ obeys $\overline{E}_\beta^\alpha = - E^\alpha_\beta$, so has sixteen degrees of freedom.   The Lie algebra commutators are easily computed directly, with the result:
\[ [E^{\alpha\beta}, E^{\gamma \delta}] = 0, \hspace{10pt}  [\overline{E}_{\alpha\beta}, \overline{E}_{\gamma \delta}] = 0,\]
\[ [ E^{\alpha}_\beta, E^{\gamma\delta}]  =   - 2  \delta_\beta^{[\gamma} E^{\delta]\alpha} , \hspace{10pt}  [E^\alpha_\beta, \overline{E}_{\gamma\delta}]  =   2  \delta^\alpha_{[\gamma} \overline{E}_{\delta]\beta}, \]  
\[ [E^{\alpha}_\beta, E^\gamma_\delta] = \delta_\beta^\gamma E^\alpha_\delta - \delta_\delta^\alpha E^\gamma_\beta, \]
\[ [ E^{\alpha\beta}, \overline{E}_{\gamma\delta}] = - 4 \delta^{[\alpha}_{[\gamma} E^{\beta]}_{\delta]}.\]
Note that the operator $E^\alpha_\beta$ generates the Lie algebra $\frak{u}(\mathbb{T})$ of $\mathbb{U}(\mathbb{T})$.  The operator $H = Z^\alpha \partial_\alpha + \overline{Z}_\alpha \overline{\partial}^\alpha$ commutes with the whole algebra of $\frak{o}(\mathbb{T})$.  Also introduce the (pure imaginary) operator $E$:
\[ E = E_\alpha^\alpha =  Z^\alpha \partial_\alpha - \overline{Z}_\alpha \overline{\partial}^\alpha.\]
Then $iE$, generates phase transformations of the twistor space: $e^{itE}$ maps $Z^\alpha$ to $e^{it} Z^\alpha$, for any real $t$.  Note that $E$ grades the algebra $\frak{o}(\mathbb{T})$:
\[ [E, E^{\alpha\beta}] = 2 E^{\alpha\beta},  \hspace{10pt} [E, E^\alpha_\beta] = 0, \hspace{10pt} [E, \overline{E}_{\alpha\beta}] = -2 \overline{E}_{\alpha\beta}.\] 
Also put $h = Z^\alpha \partial_\alpha$ and $\overline{h} = \overline{Z}_\alpha\overline{\partial}^\alpha$.  Then we have:
\[ H = h  + \overline{h}, \hspace{10pt} E = h - \overline{h}, \]
\[ h = \frac{1}{2}(H + E), \hspace{10pt} \overline{h} = \frac{1}{2}(H - E).\]
Note that all four operators $h, \overline{h}, E$ and $H$ mutually commute.  Finally note that operators of $\frak{o}(\mathbb{T})$ kill the function $Z^\alpha\overline{Z}_\alpha$ so naturally induce operators on $\mathbb{N}'$.   Also the operator $H$ preserves  $Z^\alpha\overline{Z}_\alpha$ up to a scale, so it too induces an operator on $\mathbb{N}'$.
\section{The null twistor wave operator $\square$}
We consider the Laplacian operator associated  to the pseudo-K\"ahler structure of $\mathbb{T}$, the operator $ \square = \partial_\alpha \overline{\partial}^\alpha$.  We compute the following commutator:
\[ [\square, Z^\beta\overline{Z}_\beta]  =    [\partial_\alpha \overline{\partial}^\alpha, Z^\beta\overline{Z}_\beta] = \partial_\alpha[ \overline{\partial}^\alpha, Z^\beta\overline{Z}_\beta]  + [\partial_\alpha,   Z^\beta\overline{Z}_\beta] \overline{\partial}^\alpha \]\[ = \partial_\alpha Z^\alpha + \overline{Z}_\alpha \overline{\partial}^\alpha  = 4 + Z^\alpha \partial_\alpha  + \overline{Z}_\alpha \overline{\partial}^\alpha = H + 4.\]
Then, by induction, it easily follows that for each positive integer $n$, we have the relation: 
\[  \square^n Z^\beta\overline{Z}_\beta =     Z^\beta\overline{Z}_\beta \square^n + n \square^{n-1} (H + 5 - n).\]
Now let $g(Z)$ be a given smooth function on $\mathbb{N}' $.   Extend $g(Z)$ locally to a smooth function $\tilde{g}(Z)$, defined on an open set in $\mathbb{T}$ containing the space $\mathbb{N}'$.  Then apply the operator $\square^n$ and restrict back to $\mathbb{N}'$.  Denote the result by $\square^n(g)$.  So we have:
\[ g \rightarrow \tilde{g} \rightarrow \square^n(\tilde{g})|_{\mathbb{N}'} = \square^n(g).\]
Note that the vector field $H$ is tangent to $\mathbb{N}'$, so gives a well-defined smooth vector field, still called $H$ on $\mathbb{N}'$.  For $k$ an integer,  denote by $\mathcal{H}_k$ the space of smooth functions $f(Z)$ on $\mathbb{N}'$, obeying the relation, for any $Z \in \mathbb{N}'$ and any real $t\ne 0$: 
\[ f(t Z) = t^k f(Z).\]
We use the same notation $\mathcal{H}_k$, for the associated sheaf and presheaf, the context determining which interpretation is relevant.  Suppose now that $g \in \mathcal{H}_{k}$, so $g$ obeys $H g = k g$.  Then we may consistently require that the extension $\tilde{g}$ obeys the analogous relation $H\tilde{g} = k \tilde{g}$.   Let $\tilde{g}'$ be another such extension of $g$.   Then we have $H\tilde{g}' = k \tilde{g}'$ and $\tilde{g} - \tilde{g}'$ vanishes on $\mathbb{N}'$, so by the Malgrange division theorem, we may write $\tilde{g} - \tilde{g'} =  Z^\alpha\overline{Z}_\alpha h(Z)$, where $(H - k + 2)h(Z) =  0$.  Then by our calculation above, we have:
\[ (\square^n(\tilde{g} - \tilde{g}'))(Z) =\square^n(Z^\alpha\overline{Z}_\alpha h(Z)) = Z^\alpha\overline{Z}_\alpha (\square^n(h))(Z) + n (\square^{n-1} (H + 5 - n)h)(Z)\]
\[ =  Z^\alpha\overline{Z}_\alpha (\square^n(h))(Z) + n  (k - n + 3) (\square^{n-1} h)(Z).\]
In the special case that $k = n - 3$, the last term vanishes and restricting to $\mathbb{N}'$ we get: $\square^n(\tilde{g})|_{\mathbb{N}'} =  \square^n(\tilde{g}')|_{\mathbb{N}'} $.   Hence we have $\square^n(g)$ independent of the choice of the extension of $g$ into $\mathbb{T}$, so $\square^n(g)$ is canonically defined.  We have shown:
\begin{itemize} \item The operator $\square^n$ induces naturally a map from the space (or sheaf) $\mathcal{H}_{n-3} $ to the space or sheaf $\mathcal{H}_{-3-n}$.   
\end{itemize} 
Only the case $n = 1$ will be analyzed further here.   The operator $\square$ then induces a natural second-order differential operator taking $\mathcal{H}_{-2}$ to $\mathcal{H}_{-4}$, which we call the null twistor wave operator.
\eject\noindent
\section{The null twistor wave operator constructed intrinsically in terms of the operators of the Lie algebra $\frak{o}(\mathbb{T})$}
An alternative approach to the twistor wave operator uses the operator $E^{\alpha\beta} = 2Z^{[\alpha} \overline{\partial}^{\beta]}$, which operates intrinsically to the space $\mathbb{N}'$, since it kills the quantity $Z^\alpha \overline{Z}_\alpha$. We have the following relations, using the notation of the previous section:
\[ E^{\alpha\beta} \overline{E}_{\alpha\gamma} =  4Z^{[\alpha} \overline{\partial}^{\beta]}\overline{Z}_{[\alpha} \partial_{\gamma]}\]
\[  =  Z^{\alpha} \overline{\partial}^{\beta}\overline{Z}_{\alpha} \partial_{\gamma} - Z^{\beta} \overline{\partial}^{\alpha}\overline{Z}_{\alpha} \partial_{\gamma} + Z^{\beta} \overline{\partial}^{\alpha}\overline{Z}_{\gamma} \partial_{\alpha} - Z^{\alpha} \overline{\partial}^{\beta}\overline{Z}_{\gamma} \partial_{\alpha}      \]
\[  =  Z^{\alpha} \overline{Z}_{\alpha}\overline{\partial}^{\beta} \partial_{\gamma}  - Z^\beta \partial_\gamma (\overline{h} + 2) + Z^{\beta} \overline{Z}_{\gamma}\overline{\partial}^{\alpha} \partial_{\alpha}   - ( \delta_\gamma^\beta  +\overline{Z}_{\gamma} \overline{\partial}^{\beta}) h,  \] 
\[ [ E^{\alpha\beta}, \overline{E}_{\alpha\gamma}] = - 4 \delta^{[\alpha}_{[\alpha} E^{\beta]}_{\gamma]} = - (\delta^\alpha_\alpha - 2) E^\beta_\gamma - \delta_\gamma^\beta E = - 2 E^\beta_\gamma - \delta_\gamma^\beta E, \]
\[ \frac{1}{2}(E^{\alpha\beta} \overline{E}_{\alpha\gamma} +  \overline{E}_{\alpha\gamma} E^{\alpha\beta}) = E^{\alpha\beta} \overline{E}_{\alpha\gamma}  - \frac{1}{2}[E^{\alpha\beta},  \overline{E}_{\alpha\gamma}]\]
\[ = Z^{\alpha} \overline{Z}_{\alpha}\overline{\partial}^{\beta} \partial_{\gamma}  - Z^\beta \partial_\gamma (\overline{h} + 1)   - \overline{Z}_{\gamma} \overline{\partial}^{\beta} (h + 1)  + Z^{\beta} \overline{Z}_{\gamma}\overline{\partial}^{\alpha} \partial_{\alpha} - \frac{1}{2}  \delta^{\beta}_{\gamma} (h +\overline{h}) \]
\[ = Z^{\alpha} \overline{Z}_{\alpha}\overline{\partial}^{\beta} \partial_{\gamma}  -\frac{1}{2} ( Z^\beta \partial_\gamma + \overline{Z}_{\gamma} \overline{\partial}^{\beta} + \delta^\beta_\gamma) (H + 2)  +\frac{1}{2} ( Z^\beta \partial_\gamma - \overline{Z}_{\gamma} \overline{\partial}^{\beta}) (h - \overline{h})   + Z^{\beta} \overline{Z}_{\gamma}\overline{\partial}^{\alpha} \partial_{\alpha} + \delta^{\beta}_{\gamma}. \]
\[ = Z^{\alpha} \overline{Z}_{\alpha}\overline{\partial}^{\beta} \partial_{\gamma}  -\frac{1}{2} ( Z^\beta \partial_\gamma + \overline{Z}_{\gamma} \overline{\partial}^{\beta} + \delta^\beta_\gamma) (H + 2)  +\frac{1}{2} E^\beta_\gamma E_\alpha^\alpha   + Z^{\beta} \overline{Z}_{\gamma}\overline{\partial}^{\alpha} \partial_{\alpha} + \delta^{\beta}_{\gamma}. \]
Rearranging the terms of this relation we have:
\[ \frac{1}{2}(E^{\alpha\beta} \overline{E}_{\alpha\gamma} +  \overline{E}_{\alpha\gamma} E^{\alpha\beta} - E^\beta_\gamma E_\alpha^\alpha -  2\delta^{\beta}_{\gamma} ) = Z^{\alpha} \overline{Z}_{\alpha}\overline{\partial}^{\beta} \partial_{\gamma}  -\frac{1}{2} ( Z^\beta \partial_\gamma + \overline{Z}_{\gamma} \overline{\partial}^{\beta} + \delta^\beta_\gamma) (H + 2)    + Z^{\beta} \overline{Z}_{\gamma}\overline{\partial}^{\alpha} \partial_{\alpha}. \]
In particular, when $Z^\alpha \overline{Z}_\alpha = 0$ and $H + 2 = 0$, we have just:
\[ \frac{1}{2}(E^{\alpha\beta} \overline{E}_{\alpha\gamma} +  \overline{E}_{\alpha\gamma} E^{\alpha\beta} -  E^\beta_\gamma E_\alpha^\alpha -  2\delta^{\beta}_{\gamma} ) = Z^{\beta} \overline{Z}_{\gamma}\overline{\partial}^{\alpha} \partial_{\alpha}. \]
The left-hand side of this equation is intrinsic to $\mathbb{N}'$, so therefore, so is the right-hand side, acting on $\mathcal{H}_{-2}$.
In the language of our conformal six manifold, $\mathbb{M}$, the bundles whose sections give the spaces $\mathcal{H}_k$ are conformally weighted line bundles.  In this language, we then can rephrase the result on the wave operator as follows:
\begin{itemize} \item The operator $\square$ naturally induces a conformally invariant second order wave operator on $\mathbb{M}$ from the line bundle of conformal weight $-2$ to the line bundle of conformal weight $-4$.
\end{itemize} 
\section{The co-spin bundle approach to twistor space}
In Minkowski space-time, we use complex two-component spinors, with spinor index pairs $AA'$ corresponding to vector indices $a$; conjugation interchanges primed and unprimed indices \cite{pen5}-\cite{pen1}. The metric, $g_{ab}$, and its inverse, $g^{ab}$, factorize:
\[ g_{ab} = \epsilon_{AB} \epsilon_{A'B'}, \hspace{10pt} g^{ab} = \epsilon^{AB}\epsilon^{A'B'},\hspace{10pt} \epsilon_{AB} = - \epsilon_{BA},  \hspace{10pt} \epsilon_{AB}\epsilon^{AC} = \delta_B^{\hspace{3pt}C}.\]   Here $\epsilon_{AB}$ and $\epsilon^{AB}$  are the complex conjugates of $\epsilon_{A'B'}$ and $\epsilon^{A'B'}$, respectively.  Spinor indices are raised and lowered according to the scheme $v_B = v^A \epsilon_{AB}$, $v^A = \epsilon^{AB} v_B$ and their conjugates, as appropriate.  The points of the co-spin bundle are labelled by pairs $(x^a, \pi_{B'})$, with $\pi_{B'}$ a co-spinor at the point $x^a$.  The co-spin bundle has real dimension eight and maps naturally to the co-tangent bundle via the map $(x^a, \pi_{B'}) \rightarrow (x^a, p_b)$, with $p_b = \pi_{B'} \overline{\pi}_B$.   Two points in the co-spin bundle map to the same image, with $p_a$ non-zero, if and only if their co-spinors are phase multiples of each other.   The image is the part of the cotangent bundle of space-time with $p_b$ either zero or null and future pointing.   There are natural indexed vertical vector fields $\partial^{A'} $ and $\overline{\partial}^A$, which annihilate $x^b$ and which obey the relations:
\[ \partial^{A'} \pi_{B'} = \delta_{B'}^{\hspace{5pt} A'}, \hspace{10pt}  \partial^{A'} \overline{\pi}_{B} = 0, \hspace{10pt}  \overline{\partial}^{A} \pi_{B'} = 0, \hspace{10pt}  \overline{\partial}^{A} \overline{\pi}_{B} = \delta_B^{\hspace{3pt}A}.\]
The spinor translation of the operators $h, \overline{h}, H$ and $E$ is as follows:
\[ h = \pi_{A'} \partial^{A'}, \hspace{10pt} \overline{h} = \overline{\pi}_A \overline{\partial}^A, \hspace{10pt} H = h + \overline{h} =  \pi_{A'} \partial^{A'} + \overline{\pi}_A \overline{\partial}^A, \hspace{10pt} E = h - \overline{h} =  \pi_{A'} \partial^{A'} - \overline{\pi}_A \overline{\partial}^A.\] 
$H$ generates the real scaling $\pi_{A'} \rightarrow e^t \pi_{A'}$; $iE$ generates the phase transformation $\pi_{A'} \rightarrow e^{it} \pi_{A'}$.  The space-time covariant derivative extends to the co-spin bundle such that  $\partial_a \pi_{B'} = 0$.  The null geodesic spray, denoted $\mathcal{N}$, is the vector field: 
\[ \mathcal{N} = \pi^{A'} \overline{\pi}^A \partial_a.\]
The equation $\mathcal{N} f = 0$ gives the twistor functions.  The twistor variables are:
\[ Z^\alpha = (ix^a \pi_{A'} , \pi_{A'}), \hspace{10pt} \overline{Z}_\alpha = (\overline{\pi}_A, -ix^a \overline{\pi}_{A} ).\]
Note that $\mathcal{N} Z^\alpha = 0$, $\mathcal{N} \overline{Z}_\alpha = 0$ and $Z^\alpha \overline{Z}_\alpha = 0$.   Any function killed by $\mathcal{N}$ may be regarded as a function of $Z^\alpha$.  The space of affinely parametrized null geodesics, with attached covariantly constant non-zero spinor $\pi_{A'}$ is isomorphic to the space $\mathbb{N}$ with a complex two-plane $\mathbb{I}$ removed (this two-plane corresponds to the vertex of the null cone at infinity, when Minkowski space-time is appropriately conformally compactified).   The functions $f \in \mathcal{H}_k$ are defined by the formulas:
\[ \mathcal{N} f = (H - k ) f = 0,\hspace{10pt}  f(x^a, s\pi_{A'}) = s^k f(x^a, \pi_{A'}),  \hspace{4pt} \textrm{where} \hspace{4pt} 0 \ne s \in \mathbb{R}.\]
Using the twistor variables, we have $f(sZ^\alpha) = s^k f(Z^\alpha)$, for $s$ real and non-zero.
\section{The spinor translation of the operators of $\frak{o}(\mathbb{T})$}
We write out the operators $E^{\alpha\beta}$, its conjugate $\overline{E}_{\alpha\beta}$ and $E^\alpha_\beta$ on the spin bundle.   First, for the operator $E^{\alpha\beta} = Z^\alpha \overline{\partial}^\beta  - Z^\beta \overline{\partial}^\alpha$, we have the derivatives:
\[ E^{\alpha\beta} (Z^\gamma, \overline{Z}_\gamma) = (0,    Z^\alpha\delta_\gamma^\beta  - Z^\beta\delta_\gamma^\alpha).\]
Write:
\[ E^{\alpha\beta} = A^{\alpha\beta d} \partial_d + A^{\alpha\beta}_C \overline{\partial}^C +  A^{\alpha\beta}_{C'} \partial^{C'}.\]
Then we have:
\[ 0 =  (A^{\alpha\beta d} \partial_d + A^{\alpha\beta}_{C} \overline{\partial}^C + A^{\alpha\beta}_{C'} \partial^{C'}) (ix^c\pi_{C'}, \pi_{C'}) = (i A^{\alpha\beta c}\pi_{C'} + A^{\alpha\beta}_{C'} x^c, A^{\alpha\beta}_{C'}).\]
So $A^{\alpha\beta}_{C'} =  0$ and $A^{\alpha\beta c} = \pi^{C'} X^{\alpha\beta C}$.  Then we have:
\[ \overline{\pi}_C E^{\alpha\beta} = \overline{\pi}_C  X^{\alpha\beta D}\pi^{D'} \partial_d + \overline{\pi}_C A^{\alpha\beta}_E \overline{\partial}^E  = X^{\alpha\beta}\pi^{C'} \partial_{c} + \overline{\pi}_C A^{\alpha\beta}_E \overline{\partial}^E.\]
Here $X^{\alpha\beta} = \overline{\pi}_C X^{\alpha\beta C}$.  
Finally we need:
\[  \overline{\pi}_C E^{\alpha\beta}  (- i x^d \overline{\pi}_{D}, \overline{\pi}_D) = (X^{\alpha\beta}\pi^{C'} \partial_{c} + \overline{\pi}_C A^{\alpha\beta}_E \overline{\partial}^E)  (- i x^d \overline{\pi}_{D}, \overline{\pi}_D)\]
\[ = \overline{\pi}_C (- iX^{\alpha\beta}\pi^{D'}-i A^{\alpha\beta}_D x^d,    A^{\alpha\beta}_D).\]
So we get:
\[  (- iX^{\alpha\beta}\pi^{C'}- i A^{\alpha\beta}_C x^c,    A^{\alpha\beta}_C) = 2Z^{[\alpha}\delta_\gamma^{\beta]}.\]
So we have:
\[ A^{\alpha\beta}_C =  2Z^{[\alpha}\delta_C^{\beta]},  \hspace{10pt} X^{\alpha\beta} =  2 Z^{[\alpha} X^{\beta]}, \]
\[ X^\beta \pi^{C'} + \delta_C^\beta x^c = i \delta^{C'\beta} + \lambda^{C'} Z^\beta, \]
\[ X^B \pi^{C'} + x^{BC'} =  i \lambda^{C'} x^{BB'} \pi_{B'}, \]
\[ X_{B'} \pi^{C'}  =  \lambda^{C'} \pi_{B'} + i\delta_{B'}^{C'} = \lambda_{B'} \pi^{C'} + (i+ \pi_{E'} \lambda^{E'})\delta_{B'}^{C'}  ,\]
\[ \pi_{C'} \lambda^{C'} = - i, \hspace{10pt} X_{B'} = \lambda_{B'}, \]
\[ X^B \pi^{C'} = - x^{BC'} + i \lambda^{C'} x^{BB'} \pi_{B'}  =  i\lambda_{B'} x^{BB'} \pi^{C'}, \] 
\[ X^\beta = (i x^b\lambda_{B'} , \lambda_{B'}),  \pi_{C'} \lambda^{C'} = - i.\]
\eject\noindent
So finally we may write:
\[ \overline{\pi}_C E^{\alpha\beta} = X^{\alpha\beta} \pi^{C'} \partial_c +  2\overline{\pi}_C Z^{[\alpha}\delta_E^{\beta]}\overline{\partial}^E, \hspace{10pt}  \pi_{C'} \overline{E}_{\alpha\beta} = \overline{X}_{\alpha\beta} \overline{\pi}^{C} \partial_c +  2\pi_{C'} \overline{Z}_{[\alpha}\delta_{\beta]E'}\partial^{E'}. \]
Here $X^{\alpha\beta}$ has entries:
\[ X^{AB} =  \epsilon^{AB} x_c \pi^{C'}  x^{B'C}\lambda_{B'} =  \frac{i}{2}\epsilon^{AB} x^2,\]\[X_{A'B'} = - i \epsilon_{A'B'}, \hspace{10pt} X^A_{B'} = Z^A X_{B'} - Z_{B'} X^A = x^A_{B'}.\] 
In particular, we have then the contractions:
\[ X^{\alpha\beta}\overline{X}_{\delta\beta} = 0, \hspace{10pt} X^{\alpha\beta} \overline{Z}_\beta = 0.\]
Also we have:
\[ \pi^{C'}\partial_c X^{\alpha\beta} = (-i \epsilon^{AB} \pi_{C'} x^{C'}_C,  \pi_{B'} \delta_C^A, 0) = - 2Z^{[\alpha} \delta_C^{\beta]} ,  \]
\[ X^{\alpha}_{\hspace{6pt} B'} \pi^{B'} = (- x^a\pi_{A'},  i \pi_{A'}) = i Z^\alpha, \]
\[ \partial^{E'} Z^\alpha = (i x^{AE'}, \delta_{A'}^{E'}) = \epsilon^{E'B'} (i x^{A}_{B'}, \epsilon_{A'B'})   =  i \epsilon^{E'B'} X^\alpha_{\hspace{6pt} B'},  \]
\[ \pi^{A'} \partial_a \overline{Z}_\beta = - i \overline{\pi}_A (0, \pi^{B'}).\] 
Next consider the operator:
\[ E^{\alpha}_\beta = Z^\alpha \partial_\beta - \overline{Z}_\beta \overline{\partial}^\alpha.\]
We have $E^\alpha_\beta  Z^\gamma = \delta_\beta^\gamma Z^\alpha$ and $E^\alpha_\beta  \overline{Z}_\gamma = - \delta_\gamma^\alpha \overline{Z}_\beta$.  Also we have:
\[ E^\alpha_\beta \pi_{C'} = \delta_{\beta C'} Z^\alpha, \hspace{10pt} E^\alpha_\beta \overline{\pi}_C  = - \delta_C^\alpha \overline{Z}_\beta.\]
Next we have:
\[ (E^\alpha_\beta x^c) \pi_{C'} = - i(E^\alpha_\beta Z^C) - x^c E^\alpha_\beta  \pi_{C'} = - i Z^\alpha(\delta_{\beta}^C - ix^c \delta_{\beta C'})  \]
\[ = - i (ix^{DC'}\delta_D^\alpha + \delta^{\alpha C'})(\delta_{\beta}^C - ix^{CD'} \delta_{\beta D'})\pi_{C'}.\]
\[ (E^\alpha_\beta x^c) \overline{\pi}_{C} =  i(E^\alpha_\beta \overline{Z}^{C'}) - x^c E^\alpha_\beta  \overline{\pi}_{C} = - i \overline{Z}_\beta(\delta^{\alpha C'} + ix^c \delta^{\alpha}_ {C}).  \]
So we may take:
\[ E^\alpha_\beta x^c =  - i (ix^{DC'}\delta_D^\alpha + \delta^{\alpha C'})(\delta_{\beta}^C - ix^{CD'} \delta_{\beta D'}).\]
Then we have:
\[ E^\alpha_\beta =  - i (ix^{DC'}\delta_D^\alpha + \delta^{\alpha C'})(\delta_{\beta}^C - ix^{CD'} \delta_{\beta D'})\partial_c + Z^\alpha \delta_{\beta C'} \partial^{C'} - \overline{Z}_\beta  \delta_C^\alpha \overline{\partial}^C\]
\[ =   - i X^\alpha_{E'} \overline{X}_{\delta E} \partial^e + Z^\alpha \delta_{\beta C'} \partial^{C'} - \overline{Z}_\beta  \delta_C^\alpha \overline{\partial}^C.\]
Taking the trace, we get the expected formulas:
\[ E^\alpha_\alpha =  \pi_{C'}\partial^{C'} - \overline{\pi}_C \overline{\partial}^C = h - \overline{h} = E.\]
\eject\noindent
\section{The spinor description of the null twistor wave operator}
We now are able to compute, in the spinor formalism, the following quantity, which gives the main part of the twistor wave operator:
\[ \pi_{C'} \overline{\pi}_C E^{\alpha\beta} \overline{E}_{\delta \beta} =  \overline{\pi}_C E^{\alpha\beta} \pi_{C'}\overline{E}_{\delta \beta}\]
\[ = (X^{\alpha\beta} \pi^{E'} \partial_{CE'} +  2\overline{\pi}_C Z^{[\alpha}\delta_E^{\beta]}\overline{\partial}^E )(\overline{X}_{\delta \beta} \overline{\pi}^{F} \partial_{FC'} + 2 \pi_{C'} \overline{Z}_{[\delta} \delta_{\beta]F'} \partial^{F'})\]
\[ = X^{\alpha\beta}( \pi^{E'} \partial_{CE'} \overline{X}_{\delta \beta}) \overline{\pi}^{F} \partial_{FC'} + \pi_{C'}  X^{\alpha}_{\hspace{6pt} E'}( \pi^{E'} \partial_{CE'}  \overline{Z}_{\delta})  \partial^{E'} - \pi_{C'}  X^{\alpha\beta}( \pi^{E'} \partial_{CE'}  \overline{Z}_{\beta} )\delta_{\delta E'} \partial^{E'}\]
\[ +  \pi_{C'}\overline{Z}_{\delta}X^{\alpha}_{\hspace{6pt}F'} \pi^{E'} \partial_{CE'} \partial^{F'}  +  \overline{\pi}_CZ^\alpha \overline{X}_{\delta \beta} \delta_E^{\beta}\overline{\partial}^E \overline{\pi}^{F} \partial_{FC'} +    4\overline{\pi}_C\pi_{C'} Z^{[\alpha}\delta_E^{\beta]}\overline{\partial}^E  \overline{Z}_{[\delta} \delta_{\beta]F'} \partial^{F'}\]
\[ = Z^{\alpha}\overline{X}_{\delta C} \overline{\pi}^{F} \partial_{FC'} -i \pi_{C'}\overline{\pi}_C  X^{\alpha}_{\hspace{6pt} E'}(0, \pi^{D'}) \partial^{E'} - \pi_{C'} \overline{\pi}_C Z^\alpha \delta_{\delta E'} \partial^{E'}\]
\[ +  \pi_{C'}\overline{Z}_{\delta}X^{\alpha}_{\hspace{6pt}F'} \pi^{E'} \partial_{CE'} \partial^{F'}  -  \overline{\pi}_CZ^\alpha \overline{X}_{\delta E} \partial_{C'}^E+  \overline{\pi}_C Z^\alpha \overline{X}_{\delta E}  \overline{\pi}^{F} \partial_{FC'}  \overline{\partial}^E\]\[ -    4i \overline{\pi}_C\pi_{C'} Z^{[\alpha}\delta_E^{\beta]}(\epsilon^{EF} \overline{X}_{[\delta|F|}) \delta_{\beta]F'} \partial^{F'} +   4\overline{\pi}_C\pi_{C'} Z^{[\alpha}\delta_E^{\beta]} \overline{Z}_{[\delta} \delta_{\beta]F'}\overline{\partial}^E  \partial^{F'}\]
\[ = Z^{\alpha}\overline{X}_{\delta C} \overline{\pi}^{F} \partial_{FC'} -i \pi_{C'}\overline{\pi}_C  X^{\alpha}_{\hspace{6pt} E'}(0, \pi^{D'}) \partial^{E'} - \pi_{C'} \overline{\pi}_C Z^\alpha \delta_{\delta E'} \partial^{E'}\]
\[ +  \pi_{C'}\overline{Z}_{\delta}X^{\alpha}_{\hspace{6pt}F'} \pi^{E'} \partial_{CE'} \partial^{F'}  -  \overline{\pi}_CZ^\alpha \overline{X}_{\delta E} \partial_{C'}^E+  \overline{\pi}_C Z^\alpha \overline{X}_{\delta E}  \overline{\pi}^{F} \partial_{FC'}  \overline{\partial}^E\]\[ +  i \overline{\pi}_C\pi_{C'} Z^{\alpha}(\epsilon^{EF} \overline{X}_{EF}) \delta_{\delta F'} \partial^{F'}  +    i \overline{\pi}_C\pi_{C'}\delta_E^{\alpha}(\epsilon^{EF} \overline{X}_{\delta F}) \pi_{F'} \partial^{F'}\]\[ - \overline{\pi}_C\pi_{C'} Z^{\alpha} \delta_{\delta F'}\overline{\pi}_E \overline{\partial}^E  \partial^{F'} - \overline{\pi}_C\pi_{C'} \delta_E^{\alpha} \overline{Z}_{\delta} \pi_{F'}\overline{\partial}^E  \partial^{F'}\] 
Conjugating and adding, we get:
\[  \pi_{C'} \overline{\pi}_C( E^{\alpha\beta} \overline{E}_{\delta \beta} + \overline{E}_{\delta\beta} E^{\alpha\beta}) \]
\[ = Z^{\alpha}(\overline{X}_{\delta C} \overline{\pi}^{F} \partial_{FC'}  - \pi_{C'} \overline{\pi}_C \delta_{\delta E'} \partial^{E'} + \overline{\pi}_C \overline{X}_{\delta F} \overline{\pi}^E \partial_{EC'} \overline{\partial}^F -  \overline{\pi}_C \overline{X}_{\delta E} \partial_{C'}^E+  \overline{\pi}_C \overline{X}_{\delta E}  \overline{\pi}^{F} \partial_{FC'}  \overline{\partial}^E\]
 \[ - 2 \overline{\pi}_C\pi_{C'}  \delta_{\delta F'} \partial^{F'}    - \overline{\pi}_C\pi_{C'}  \delta_{\delta F'}\overline{\pi}_E \overline{\partial}^E  \partial^{F'} - \overline{\pi}_C\pi_{C'} \delta_{E' \delta} \overline{\pi}_{F}\partial^{E'}  \overline{\partial}^{F})\]  \[ -i \pi_{C'}\overline{\pi}_C(  X^{\alpha}_{\hspace{6pt} F'}\delta_{\delta E'}  \pi^{E'}  - \delta_E^{\alpha}(\epsilon^{EF} \overline{X}_{\delta F}) \pi_{F'}) \partial^{F'}+ c.c.\]
 Now we have:
 \[ X^\alpha_{B'} = (X^A_{B'} , X_{A'B'}) = (x^A_{B'}, - i\epsilon_{A'B'})  \]\[ = - i \epsilon_{C'B'}( ix^{AC'}, \delta_{A'}^{C'}) = - i \epsilon_{C'B'}(i x^c\delta_C^\alpha +\delta^{\alpha C'}).\]
  \eject\noindent
So we get:
\[  X^{\alpha}_{\hspace{6pt} F'}\delta_{\delta E'}  \pi^{E'}  - \delta_E^{\alpha}(\epsilon^{EF} \overline{X}_{\delta F}) \pi_{F'}= X^{\alpha}_{\hspace{6pt} F'}\delta_{\delta E'}  \pi^{E'}  - i\delta_E^{\alpha}(\epsilon^{EF}  \epsilon_{CF}(- i x^c\delta_{C'\delta} +\delta^{ C}_\delta) \pi_{F'}\]
\[ = X^{\alpha}_{\hspace{6pt} F'}\delta_{\delta E'}  \pi^{E'}  - i\delta_C^{\alpha}(- i x^c\delta_{C'\delta} +\delta^{ C}_\delta) \pi_{F'}\]
\[ =  \epsilon^{E'B'}X^{\alpha}_{B'}\delta_{\delta E'}  \pi_{F'} +  X^{\alpha}_{G'}\pi^{G'} \delta_{\delta F'}   - i\delta_C^{\alpha}(- i x^c\delta_{C'\delta} +\delta^{ C}_\delta) \pi_{F'}\] 
\[ = - i(i x^c\delta_C^\alpha +\delta^{\alpha C'})\delta_{\delta C'}  \pi_{F'} +  iZ^\alpha \delta_{\delta F'}   - i\delta_C^{\alpha}(- i x^c\delta_{C'\delta} +\delta^{ C}_\delta) \pi_{F'}\] 
\[ = - i \delta^{\alpha}_\delta  \pi_{F'} +  iZ^\alpha \delta_{\delta F'}.\]
So we now have the formula, using the relations $h = \pi_{A'} \partial^{A'}$, $\overline{h} = \overline{\pi}_A \overline{\partial}^A$, $H = h + \overline{h}$ and $E = h - \overline{h}$:
\[   \pi_{C'} \overline{\pi}_C( (\delta^\alpha_\delta +  Z^\alpha \delta_{\delta E'} \partial^{E'} +  \overline{Z}_\delta \delta^\alpha_E \overline{\partial}^E) (H+ 2) + E^{\alpha\beta} E_{\delta \beta} + E_{\delta\beta} E^{\alpha\beta} - 2\delta_\delta^\alpha -( Z^\alpha \delta_{\delta E'} \partial^{E'} - \overline{Z}_\delta \delta^\alpha_E \overline{\partial}^E)E) \]
\[ = Z^{\alpha}( \overline{X}_{\delta C} \overline{\pi}^{F} \partial_{FC'}  -  \overline{\pi}_C \overline{X}_{\delta E} \partial_{C'}^E + 2\overline{\pi}_C \overline{X}_{\delta F} \overline{\pi}^E \partial_{EC'} \overline{\partial}^F) + c.c.\] 
\[ = Z^{\alpha}( \overline{X}_{\delta C} \overline{\pi}^{F} \partial_{FC'}  -  \overline{\pi}_C \overline{X}_{\delta E} \partial_{C'}^E + 2\overline{\pi}_C \epsilon^{EG} \overline{X}_{\delta G} \overline{\pi}_F \partial_{EC'} \overline{\partial}^F +  2\overline{\pi}_C \overline{X}_{\delta D} \overline{\pi}^D \partial_{FC'} \overline{\partial}^F) + c.c.\] 
\[ = Z^{\alpha}( - \overline{X}_{\delta C} \overline{\pi}_{E} \partial_{C'}^E  +  \overline{\pi}_C \overline{X}_{\delta E} \partial_{C'}^E + \overline{\pi}_C \epsilon^{EG} \overline{X}_{\delta G} \partial_{EC'} (H - E + 2) - 2i \overline{Z}_\delta \overline{\pi}_C  \partial_{FC'} \overline{\partial}^F) + c.c.\] 
\[ = Z^{\alpha}( \overline{X}_{\delta E} \overline{\pi}^{E} \partial_{CC'} + \overline{\pi}_C \epsilon^{EG} \overline{X}_{\delta G} \partial_{EC'} (H - E + 2) - 2i \overline{Z}_\delta \overline{\pi}_C  \partial_{FC'} \overline{\partial}^F) + c.c.\] 
\[ = Z^{\alpha}(  \overline{\pi}_C \epsilon^{EG} \overline{X}_{\delta G} \partial_{EC'} (H - E + 2) - 2i \overline{Z}_\delta \overline{\pi}_C  \partial_{FC'} \overline{\partial}^F) + c.c.\] 
We re-write the $- E$-coefficient of the right-hand side of this equation; this coefficient is:
\[  Y^\alpha_\delta = Z^{\alpha} \overline{\pi}_C \epsilon^{EG} \overline{X}_{\delta G} \partial_{EC'}  - c.c.\]
\[ =  i\overline{\pi}_C X^\alpha_{E'} \pi^{E'}  \overline{X}_{\delta G} \partial_{C'}^G - c.c.\]
\[ =  i \pi_{C'} \overline{\pi}_C X^\alpha_{E'}  \overline{X}_{\delta G} \partial^{GE'} +  i X^\alpha_{C'} \pi^{D'} \overline{\pi}_C \overline{X}_{\delta G} \partial_{D'}^G  - c.c. \] 
\[ =  i \pi_{C'} \overline{\pi}_C X^\alpha_{E'}  \overline{X}_{\delta G} \partial^{GE'} +  i X^\alpha_{C'} \pi^{D'} \overline{\pi}^G \overline{X}_{\delta G} \partial_{D'C} -  i X^\alpha_{C'} \overline{X}_{\delta C}\pi^{D'} \overline{\pi}^D \partial_{D'D}  - c.c. \] 
\[ =  i \pi_{C'} \overline{\pi}_C X^\alpha_{E'} \overline{X}_{\delta G} \partial^{GE'} +  \overline{Z}_\delta X^\alpha_{C'} \pi^{D'} \partial_{D'C} -  i X^\alpha_{C'} \overline{X}_{\delta C}\pi^{D'} \overline{\pi}^D \partial_{D'D}  - c.c. \] 
\[ =  i \pi_{C'} \overline{\pi}_C X^\alpha_{E'}  \overline{X}_{\delta G} \partial^{GE'} -  \overline{Z}_\delta X^\alpha_{D'} \pi_{C'} \partial_{C}^{D'} +   \overline{Z}_\delta X^\alpha_{E'} \pi^{E'} \partial_{C'C} -  i X^\alpha_{C'} \overline{X}_{\delta C}\pi^{D'} \overline{\pi}^D \partial_{D'D}  - c.c. \] 
\[ =  i \pi_{C'} \overline{\pi}_C X^\alpha_{E'}  \overline{X}_{\delta G} \partial^{GE'} -  \overline{Z}_\delta X^\alpha_{D'} \pi_{C'} \partial_{C}^{D'} +   iZ^\alpha \overline{Z}_\delta \partial_{C'C} -  i X^\alpha_{C'} \overline{X}_{\delta C}\pi^{D'} \overline{\pi}^D \partial_{D'D}  - c.c. \] 
\[ =   i \pi_{C'} \overline{\pi}_C X^\alpha_{E'}  \overline{X}_{\delta G} \partial^{GE'} -  \overline{Z}_\delta X^\alpha_{D'} \pi_{C'} \partial_{C}^{D'} +   iZ^\alpha \overline{Z}_\delta \partial_{C'C} -  i X^\alpha_{C'} \overline{X}_{\delta C}\pi^{D'} \overline{\pi}^D \partial_{D'D}  - c.c. \] 
\[ = - Y_\delta^\alpha +   2 i \pi_{C'} \overline{\pi}_C X^\alpha_{E'}  \overline{X}_{\delta G} \partial^{GE'}  +   2iZ^\alpha \overline{Z}_\delta \partial_{C'C} - 2 i X^\alpha_{C'} \overline{X}_{\delta C}\pi^{D'} \overline{\pi}^D \partial_{D'D}  \] 
\[ =   i \pi_{C'} \overline{\pi}_C X^\alpha_{E'}  \overline{X}_{\delta E} \partial^{EE'}  +   iZ^\alpha \overline{Z}_\delta \partial_{C'C} -  i X^\alpha_{C'} \overline{X}_{\delta C}\pi^{D'} \overline{\pi}^D \partial_{D'D}  \] 
Now assume that $H + 2 = 0$ and $\mathcal{N} = \pi^{D'} \overline{\pi}^D \partial_{D'D}   = 0$.  Also note that the operators $\mathcal{N}$ and $E$ commute.   Then we have:
\[  \pi_{C'} \overline{\pi}_C(  E^{\alpha\beta} E_{\delta \beta} + E_{\delta\beta} E^{\alpha\beta} - 2\delta_\delta^\alpha -( - i X^\alpha_{D'}\overline{X}_{\delta D} \partial^{DD'} + Z^\alpha \delta_{\delta E'} \partial^{E'} - \overline{Z}_\delta \delta^\alpha_E \overline{\partial}^E)E)   \] 
\[ =  - i Z^{\alpha}\overline{Z}_\delta( 2\overline{\pi}_C  \partial_{FC'} \overline{\partial}^F - 2\pi_{C'} \partial_{F'C} \partial^{F'} + \partial_{CC'} E) .\] 
From the work of the last section, we recognize the term $  - i X^\alpha_{D'}\overline{X}_{\delta D} \partial^{DD'} + Z^\alpha \delta_{\delta E'} \partial^{E'} - \overline{Z}_\delta \delta^\alpha_E \overline{\partial}^E$ as being exactly the operator $E^\alpha_\delta$.   So,  substituting, we have, 
provided $H + 2 = \mathcal{N}  = 0$: 
\[  \pi_{C'} \overline{\pi}_C(  E^{\alpha\beta} E_{\delta \beta} + E_{\delta\beta} E^{\alpha\beta} - 2\delta_\delta^\alpha - E^\alpha_\delta E)   \] 
\[ =  - i Z^{\alpha}\overline{Z}_\delta( 2\overline{\pi}_C  \partial_{FC'} \overline{\partial}^F - 2\pi_{C'} \partial_{F'C} \partial^{F'} + \partial_{CC'} E) .\]
The left-hand side of this equation is exactly $2\pi_{C'} \overline{\pi}_C$ times the twistor wave operator.  
We have proved that the spinor description of the operator $\square$, induced by $\partial_\alpha \overline{\partial}^\alpha$, on twistor functions $f$, of total degree minus two, is given by the formula:
\[  \pi_{C'} \overline{\pi}_C \hspace{1.5pt} \square (f) =    i \left(\pi_{C'}  \partial^{F'} \partial_{F'C}f - \overline{\pi}_C \overline{\partial}^F  \partial_{FC'} \overline{\partial}^Ff - \frac{1}{2} \partial_{CC'} Ef\right) .\]
It is not immediately obvious that the right-hand side of this equation is proportional to $\pi_{C'} \pi_C$.   To verify this directly, first note that since $\mathcal{N} f = 0$, we may write $\partial_a f = \pi_{A'} f_A + \overline{\pi}_A \overline{f}_{A'}$, for some $f_A$.     Then we have:
\[ \pi_{C'}  \partial^{F'} \partial_{F'C}f  - \overline{\pi}_C \overline{\partial}^F  \partial_{FC'}f  - \frac{1}{2} \partial_{CC'} E f \]
\[ = \pi_{C'} (h + 2) f_C - \overline{\pi}_C (\overline{h} + 2) \overline{f}_{C'} + \pi_{C'} \overline{\pi}_C( \partial^{F'} \overline{f}_{F'} - \overline{\partial}^F f_F)  - \frac{1}{2}E \partial_{CC'}  f \]
\[  = \frac{1}{2}((2h + 2 - E)(\pi_{C'} f_C) - (2\overline{h} + 2 + E)( \overline{\pi}_C  \overline{f}_{C'}) ) + \pi_{C'} \overline{\pi}_C( \partial^{F'}  \overline{f}_{F'} - \overline{\partial}^F f_F)  \]
\[  =\frac{1}{2}  (H + 2)( \pi_{C'} f_C -  \overline{\pi}_C  \overline{f}_{C'}) + \pi_{C'} \overline{\pi}_C( \partial^{F'}  \overline{f}_{F'} - \overline{\partial}^F f_F). \]
It remains to show that the quantity $h_c = (H + 2)( \pi_{C'} f_C -  \overline{\pi}_C  \overline{f}_{C'}) $ is proportional to $\pi_{C'} \overline{\pi}_C$.   Since the co-vector $h_{c}$ is purely imaginary, it suffices to show that $\pi^{C'} h_{c} = 0$.  We have:
\[ \pi^{C'} h_{c}  =  \pi^{C'} (H + 2)( \pi_{C'} f_C -  \overline{\pi}_C  \overline{f}_{C'})\]\[ =  (H + 1)(\pi^{C'} (\pi_{C'} f_C -  \overline{\pi}_C  \overline{f}_{C'})) = (H + 1)(\pi^{C'} (- \pi_{C'} f_C -  \overline{\pi}_C  \overline{f}_{C'}))\] 
\[ = - (H + 1)(\pi^{C'} \partial_c f)  = - \pi^{C'} (H + 2) \partial_c f  = -  \pi^{C'}\partial_c (H + 2) f = 0.\]
Here we used that $f$ has total degree minus two and that the operators $\partial_a$ and $H$ commute, whereas $H\pi_{A'} = \pi_{A'}(H + 1)$.  
\eject\noindent
Next we verify  that the quantity   $ i (\pi_{C'}  \partial^{F'} \partial_{F'C}f - \overline{\pi}_C \overline{\partial}^F  \partial_{FC'} \overline{\partial}^Ff - \frac{1}{2} \partial_{CC'} Ef)$ is killed by the operator $\mathcal{N}$ and is of total degree $-2$.   That the degree is $-2$ is immediate, since $f$ is of degree minus two and since $H$ commutes with the operators  $E$, $\partial_a$, $\pi_{A'} \partial^{B'} $ and $\overline{\pi}_A \overline{\partial}^{B}$.  It follows that $\square(f)$ has total degree $-4$, as expected.  Then we have, since $\mathcal{N}$ commutes with $\partial_a$ and with $E$ and kills $f$:
\[ \pi^{D'} \overline{\pi}^D \partial_d( i (\pi_{C'}  \partial^{F'} \partial_{F'C}f - \overline{\pi}_C \overline{\partial}^F  \partial_{FC'} \overline{\partial}^Ff - \frac{1}{2} \partial_{CC'} Ef))
\]
\[ =  - i (\pi_{C'}   \overline{\pi}_D \partial^d \partial_{D'C}f - \overline{\pi}_C \pi_{D'} \partial^d \partial_{DC'} \overline{\partial}^Ff) = - \frac{i}{2} \pi_{C'}\pi_C( \partial_a \partial^a f - \partial_a\partial^a f) = 0. \]
Since $\mathcal{N}$ also commutes with $\pi_{C'} \overline{\pi}_C$, we have shown that the function $\square(f)$ lies in $\mathcal{H}_{-4}$, as expected.  We introduce the Lorentz generators of the co-spin-bundle:
\[ \partial_{A'}^{B'} =  \pi_{A'} \partial^{B'}  - \frac{1}{2} \pi_{C'} \partial^{C'}.\]
Note that $\partial_{A'}^{A'} = 0$, so $\partial_{A'B'} = \partial_{B'A'}$.   Then we may write our operator on $\mathcal{H}_{-2}$ as:
\[  \pi_{C'} \overline{\pi}_C\hspace{1.5pt}\square=    i (\partial_{C'}^{F'} \partial_{F'C} - \overline{\partial}_C^F \partial_{FC'}) .\]
Finally, let $g_{A}$ and $h_{A'}$ be any smooth functions, not necessarily complex conjugates of each other, such that $\partial_a f = \pi_{A'} g_A + \overline{\pi}_A h_{A'}$.  So we have $g_A = f_{A} + \overline{\pi}_{A}p$ and $h_{A'} = \overline{f}_{A'} - \pi_{A'} p$, for some smooth function $p$.  Then we have:
\[ \frac{1}{2}  (H + 2)( \pi_{C'} g_C -  \overline{\pi}_C  \overline{h}_{C'}) + \pi_{C'} \overline{\pi}_C( \partial^{F'}  \overline{h}_{F'} - \overline{\partial}^F g_F) \]
\[ = - i \pi_{C'} \overline{\pi}_C \hspace{1.5pt}\square(f) +   (H + 2)( \pi_{C'} \pi_C  p)  - \pi_{C'} \overline{\pi}_C( \partial^{F'} ( \pi_{F'} p ) + \overline{\partial}^F (\pi_F p))  \]
\[ = \pi_{C'} \overline{\pi}_C (- i \square(f) +   (H + 4)p  - (h + 2) p - ( \overline{h} + 2) p) = - i \pi_{C'} \pi_C \square(f).  \]
We have proved that the twistor wave operator $\square$ maps $\mathcal{H}_{-2}$ to $\mathcal{H}_{-4}$.   Acting on a function $f(x, \pi)$, which obeys $\mathcal{N} f = (H + 2)f = 0$ and $f(x, - \pi) = f(x, \pi)$, we have:
\[ \pi_{A'} \overline{\pi}_A \hspace{1.5pt}\square(f) =  i (\partial_{A'}^{B'} \partial_{B'A} - \overline{\partial}_A^B \partial_{BA'}) .\]
Alternatively, we first write $\partial_a f = \pi_{A'} f_A + \overline{\pi}_A f_{A'}$, with $f_A$ and $f_{A'}$ smooth.   Then:
\[ - i\pi_{A'} \overline{\pi}_A\hspace{1.5pt} \square(f) =\frac{1}{2}  (H + 2)( \pi_{A'} f_A -  \overline{\pi}_A  f_{A'}) + \pi_{A'} \overline{\pi}_A( \partial^{B'} f_{B'} - \overline{\partial}^B f_B). \]
In particular, if we arrange that $(H+ 3)f_A = 0$ and $(H + 3)f_{A'} = 0$ (or indeed, if we arrange just one of these relations, since the other then follows automatically), then we have just:
\[  \partial_a f = \pi_{A'} f_A + \overline{\pi}_A f_{A'}, \hspace{10pt} (H+ 3)f_A = 0, \hspace{10pt} (H+ 3)f_{A'} = 0, \hspace{10pt}  \square(f) = i(\partial^{B'} f_{B'} - \overline{\partial}^B f_B). \]
\eject\noindent
\section{The spinor integral operator: the $\Xi$-transform}
We next construct a spinor integral operator which goes back from $\mathcal{H}_{-4}$ to $\mathcal{H}_{-2}$.  Let $f(x, \pi) $ be a twistor function of total degree $-4$ (i.e. $f \in \mathcal{H}_{-4}$), so we have: 
\[ 0 = \pi^{A'} \overline{\pi}^{A}\partial_a f(x, \pi) = (\pi_{A'} \partial^{A'} + \overline{\pi}_A \overline{\partial}^A  + 4)f(x, \pi), \hspace{10pt} f(x, - \pi) = f(x, \pi).\]
Then we define the $\Xi$-transform $\Xi(f)(x, \eta)$ of $f$ by the formula:
\[ \Xi(f)(x, \eta) = i\int_{\pi_{A'} \eta^{A'} > 0}  f(x^a + s\overline{\eta}^A \eta^{A'}, \pi_{C'}) ds \pi^{B'} d\pi_{B'} \overline{\pi}^B d\overline{\pi}_B.\] 
Here the variable $s$ ranges over the whole real line.  Also the spinor integral is taken in the space of spinors $\pi_{A'}$ such that $\pi_{A'} \eta^{A'} $ is real and positive.   We assume that $f$ is sufficiently smooth and well-behaved at infinity such that all the integrals under consideration converge nicely. Note that the requirement that $f(x, \pi)$ be of degree minus four is natural, since the differential form $\pi^{B'} d\pi_{B'} \overline{\pi}^B d\overline{\pi}_B$ has weight four, so the total weight of the integrand  is zero, as required for the integral to be well-defined.  If we write the transform out with the symplectic spinor forms written in explicitly, we have:
\[ \Xi(f)(x, \eta) = i \int_{\pi_{A'} \eta_{B'}\epsilon^{A'B'} > 0}  f(x^a + s \epsilon^{AB} \epsilon^{A'B'}\overline{\eta}_B \eta_{B'}, \pi_{C'})  ds \epsilon^{D'E'} \epsilon^{DE}\pi_{D'}  d\pi_{E'}  \overline{\pi}_D d\overline{\pi}_E.\] 
The point here now is that under a real conformal transformation $\epsilon^{AB} \rightarrow u \epsilon^{AB}$ with $u > 0$ real, combined with the replacement $s \rightarrow u^{-2}s$ the integral is invariant.  So the integral is conformally invariant.\\\\
We need to check that the differential form being integrated is closed.    So we need to show that if $f(\pi_{A'})$ is smooth and obeys the relation: $0 = (\pi_{A'}\partial^{A'} + \overline{\pi}_{A}\overline{\partial}^{A} + 4)f $, then the following differential three-form vanishes identically on the space where $\pi_{A'} \eta^{A'} $ is real and $\pi_{A'}$ is non-zero, for any fixed non-zero spinor $\eta^{A'}$:
\[ d( f  \pi^{B'} d\pi_{B'} \overline{\pi}^B d\overline{\pi}_B )  \]\[ =  (d\pi_{A'}\partial^{A'}f + d\overline{\pi}_{A}\overline{\partial}^{A}f)  \pi^{B'} d\pi_{B'} \overline{\pi}^B d\overline{\pi}_B +  fd\pi^{B'} d\pi_{B'} \overline{\pi}^B d\overline{\pi}_B - f\pi^{B'} d\pi_{B'} d\overline{\pi}^B d\overline{\pi}_B\]
\[ =\frac{1}{2} ((\pi_{A'}\partial^{A'} + 2)f)  d\pi^{B'} d\pi_{B'} \overline{\pi}^B d\overline{\pi}_B  - \frac{1}{2} ((\overline{\pi}_{A}\overline{\partial}^{A} + 2)f)\pi^{B'} d\pi_{B'} d\overline{\pi}^B d\overline{\pi}_B\]
\[ =\frac{1}{2}( (\pi_{A'}\partial^{A'} + 2)f ) \xi, \]\[  \xi = d\pi^{B'} d\pi_{B'} \overline{\pi}^B d\overline{\pi}_B + \pi^{B'} d\pi_{B'} d\overline{\pi}^B d\overline{\pi}_B.\]
\eject\noindent
So it remains to be shown that the differential form $\xi$ vanishes identically.  To do this efficiently, first note that since $\pi_{A'} \eta^{A'}$ is real, by assumption, for each fixed $\eta_{A'}$, the space of allowable spinors $\pi_{A'}$ lies in a real vector space of three real dimensions.  So the four-form $\Xi = d\xi = 2d\pi^{B'} d\pi_{B'} d\overline{\pi}^B d\overline{\pi}_B$ vanishes identically.  Also the vector field $H = \pi_{A'} \partial^{A'} + \overline{\pi}_{A}\overline{\partial}^A  $ is tangent to the relation obeyed by $\pi_{A'}$: $0 =  \pi_{A'} \eta^{A'}- \overline{\pi}_A \overline{\eta}^A$.   Contracting $\Xi$ with $H$ gives the form $4\xi$.   Since the form $\Xi$ vanishes, so does any contraction of $\Xi$, so $\xi$ vanishes also, as required and we are done.
\\\\Note that using the transformation $(\pi_{A'}, \eta_{B'}) \rightarrow (-\pi_{A'} , - \eta_{B'})$, we find that $\Xi(f)(x, \eta) = \Xi(f)(x, -\eta)$.  Next note that under the replacement $x^a \rightarrow x^a + k \overline{\eta}^A\eta^{A'}$, combined with the variable replacement $s \rightarrow s - k$, the integral remains invariant, so $\Xi(f)(x, \eta)$ is constant along the null geodesic spray, so is a twistor function:
\[ \eta^{A'} \overline{\eta}^{A}\partial_a f(x, \eta) = 0.\]
Next if we scale $\eta^{A'} \rightarrow p \eta^{A'}$, where $p$ is real and positive, combined with a variable change $s\rightarrow sp^{-2}$, then the integral scales as $\Xi(f)(x, p \eta) = p^{-2} \Xi(f)(x, \eta)$.   So we have proved the desired relation:
\begin{itemize} \item The transform $f \rightarrow \Xi(f)$ maps the space $\mathcal{H}_{-4}$ to the space $\mathcal{H}_{-2}$.
\end{itemize}
We call this transform the $\Xi$ transform.  We can rewrite the $\Xi$-transform to obviate the requirement that $\pi_{A'} \eta^{A'}$ be real as follows:
\[ \Xi(f)(x, \eta) = i\int f\left(x^a + s\overline{\eta}^A \eta^{A'},\frac{ \pi_{C'}}{\pi_{E'} \eta^{E'}}\right) ds\frac{ \pi^{B'} d\pi_{B'} \overline{\pi}^B d\overline{\pi}_B}{(\pi_{E'} \eta^{E'}\overline{\pi}_E \overline{\eta}^E)^2}.\] 
Here we have used the fact that $f$ is of homogeneous of degree minus four in the variable $\pi_{A'}$ under positive real scalings.  The point now is that the integrand is invariant under complex scalings $\pi_{A'} \rightarrow s\pi_{A'}$ with $s$ any non-zero complex number, so in the integral we no longer need to constrain $\pi_{A'}$ by the requirement that $\pi_{A'}\eta^{A'}$ be real.  Note, however, the key fact that the integral is not in general invariant under phase transformations $\eta_{A'} \rightarrow \lambda \eta_{A'}$ with $|\lambda| = 1$.   The integral may now be construed as taken over $\mathbb{R} \times \mathbb{S}^2$, where the $\mathbb{R}$-factor corresponds to the $s$ variable and the $\mathbb{S}^2$-factor represents the Riemann sphere corresponding to the complex projective space of the co-spin space with its origin deleted.     Since this formula is homogeneous in the variable $\pi_{A'}$, we can also normalize the spinor $\pi_{A'}$ by the relation $\pi_{A'} \eta^{A'} = 1$ and then the transform formula reads:
\[ \Xi(f)(x, \eta) = i\int_{\pi_{A'} \eta^{A'} = 1}  f(x^a + s\overline{\eta}^A \eta^{A'}, \pi_{C'}) ds \pi^{B'} d\pi_{B'} \overline{\pi}^B d\overline{\pi}_B.\] 
\eject \noindent 
\section{The equation $\square \circ \Xi = 0$}
We begin by re-writing the basic $\Xi$-transform formula, using the fact that $f(x, \pi)$ is a twistor function, so may be written as $f(\omega^A, \pi_{A'})$, where $\omega^A = ix^a \pi_{A'}$:
\[ \Xi(f)(x, \eta) = i\int f\left(\frac{ ix^a \pi_{A'}}{\pi_{E'} \eta^{E'}} + is\overline{\eta}^A,\frac{ \pi_{C'}}{\pi_{E'} \eta^{E'}}\right) ds\frac{ \pi^{B'} d\pi_{B'} \overline{\pi}^B d\overline{\pi}_B}{(\pi_{E'} \eta^{E'} \overline{\pi}_E \overline{\eta}^E)^2}.\] 
Note that, since $f \in \mathcal{H}_{-4}$, the function $f(\omega^A, \pi_{A'})$ obeys the scaling relation $f(t \omega^A, t \pi_{A'}) = t^{-4} f( \omega^A, \pi_{A'})$, for any positive real number $t$.  Inside the integral, denote the operator  $\displaystyle{\frac{\partial}{\partial \omega^A}}$ by $\partial_A$, with conjugate $\overline{\partial}_{A'}$.  Note that the quantity $\partial_A f $ obeys the scaling relation $(\partial_A f)(t \omega^A, t \pi_{A'}) = t^{-5} f( \omega^A, \pi_{A'})$, for any positive real number $t$.
Applying the derivative $\partial_a$ to the integral formula, inside the integral we have:
\[ \partial_a \Xi(f)(x, \eta)  =   - \int  \left(  \left(\frac{\pi_{A'}}{\pi_{E'} \eta^{E'}}\partial_A   - \frac{\overline{\pi}_{A}}{ \overline{\pi}_{E} \overline{\eta}^{E} }\overline{\partial}_{A'}\right) f\right)  \left(\frac{ ix^a \pi_{A'}}{\pi_{E'} \eta^{E'}} + is\overline{\eta}^A,\frac{ \pi_{C'}}{\pi_{E'} \eta^{E'}}\right)  \frac{ds \pi^{B'} d\pi_{B'} \overline{\pi}^B d\overline{\pi}_B}{(\pi_{E'} \eta^{E'} \overline{\pi}_E \overline{\eta}^E)^2} \] 
We re-write the operator part of the integrand, acting on the function $\displaystyle{f \left(\frac{ ix^a \pi_{A'}}{\pi_{E'} \eta^{E'}} + is\overline{\eta}^A,\frac{ \pi_{C'}}{\pi_{E'} \eta^{E'}}\right)}$ as follows:
\[ -  \frac{1}{(\pi_{E'} \eta^{E'}\overline{\pi}_E\overline{\eta}^E)^2} \left(\frac{\pi_{A'}}{\pi_{E'} \eta^{E'}}\partial_A   - \frac{\overline{\pi}_{A}}{ \overline{\pi}_{E} \overline{\eta}^{E} }\overline{\partial}_{A'}\right)  =   -  \frac{1}{(\pi_{E'} \eta^{E'}\overline{\pi}_E\overline{\eta}^E)^3}( \overline{\pi}_B\overline{\eta}^B\pi_{A'}\partial_A - \pi_{B'}\eta^{B'} \overline{\pi}_A \overline{\partial}_A ) \]\[ =  \frac{1}{(\pi_{E'} \eta^{E'}\overline{\pi}_E\overline{\eta}^E)^3}   (\pi_{A'}\overline{\pi}^{B} \overline{\eta}_A\partial_B - \overline{\pi}_{A}\pi^{B'} \eta_{A'}\overline{\partial}_{B'} -    \pi_{A'}\overline{\pi}_{A}( \overline{\eta}^{C}\partial_C -  \eta^{C'}\overline{\partial}_{C'}))\]
 \[ =  \frac{1}{(\pi_{E'} \eta^{E'}\overline{\pi}_{E} \overline{\eta}^{E})^3 }\left( \pi_{A'}\overline{\pi}^{B} \overline{\eta}_A\partial_B - \overline{\pi}_{A}\pi^{B'} \eta_{A'}\overline{\partial}_{B'} +  i \pi_{A'}\overline{\pi}_{A}\frac{\partial}{\partial s}\right) .\]
Integrating out the $\displaystyle{\frac{\partial}{\partial s}}$ term, we get the relation:
\[ \partial_a \Xi(f)(x, \eta)   =  \eta_{A'} \overline{f}_A(x, \eta) + \overline{\eta}_A f_{A'}(x, \eta), \]
\[ f_{A'}(x, \eta)  =     \int \pi_{A'}\overline{\pi}^{B} (\partial_B  f) \left(\frac{ ix^a \pi_{A'}}{\pi_{E'} \eta^{E'}} + is\overline{\eta}^A,\frac{ \pi_{C'}}{\pi_{E'} \eta^{E'}}\right)   \frac{ds \pi^{B'} d\pi_{B'} \overline{\pi}^B d\overline{\pi}_B}{(\pi_{E'} \eta^{E'}\pi_E \eta^E)^3}. \]
Note that under the replacement $\eta_{A'} \rightarrow - \eta_{A'}$, we get $f_{A'}(x, - \eta) = - f_{A'}(x, \eta)$.
\eject\noindent
Next note that under the scaling transformation $\eta_{A'} \rightarrow t \eta_{A'} $, with $t > 0$, combined with the transformation $s \rightarrow t^{-2} s$,  the function $\displaystyle{(\partial_B  f )\left(\frac{ ix^a \pi_{A'}}{\pi_{E'} \eta^{E'}} + is\overline{\eta}^A,\frac{ \pi_{C'}}{\pi_{E'} \eta^{E'}}\right)  }$ scales by a factor of $t^5$.  Also the differential form  $\displaystyle{\frac{ds \pi^{B'} d\pi_{B'} \overline{\pi}^B d\overline{\pi}_B}{(\pi_{E'} \eta^{E'}\pi_E \eta^E)^3}}$ scales by a factor of $t^{-8}$.   So the integral scales by a factor of $t^{-3}$ and we have the relation:
\[ f(x, t\eta) = t^{-3} f(x, \eta), \hspace{4pt} \textrm{for non-zero real number} \hspace{4pt} t.\]    Denote the derivative operator $\displaystyle{\frac{\partial}{\partial \eta_{A'}}}$ by $D^{A'}$, with conjugate $\overline{D}^A$.  Then the quantity $f_{A'}(x, \eta)$ obeys the homogeneity relation: 
\[ (\eta_{B'} D^{B'} + \overline{\eta}_B \overline{D}^B + 3) f_{A'} = 0.\] 
Next we have, using the fact that $\pi_{A'} D^{A'} $ annihilates both the quantities $\pi_{A'}\eta^{A'}$ and $\overline{\pi}_A \overline{\eta}^A$:
\[ D^{A'} f_{A'} =   \int  \pi_{A'}D^{A'} \overline{\pi}^{B}( \partial_B  f) \left(\frac{ ix^a \pi_{A'}}{\pi_{E'} \eta^{E'}} + is\overline{\eta}^A,\frac{ \pi_{C'}}{\pi_{E'} \eta^{E'}}\right)   \frac{ds \pi^{B'} d\pi_{B'} \overline{\pi}^B d\overline{\pi}_B}{(\pi_{E'} \eta^{E'}\pi_E \eta^E)^3} \]
\[    =  \int  - is \pi^{B'} \overline{\pi}^{B}(\overline{\partial}_{B'}  \partial_B  f )\left(\frac{ ix^a \pi_{A'}}{\pi_{E'} \eta^{E'}} + is\overline{\eta}^A,\frac{ \pi_{C'}}{\pi_{E'} \eta^{E'}}\right)   \frac{ds \pi^{B'} d\pi_{B'} \overline{\pi}^B d\overline{\pi}_B}{(\pi_{E'} \eta^{E'}\pi_E \eta^E)^3}.\]
The right-hand side of this equation  is real, so immediately we have the differential equation:
\[ D^{A'} f_{A'} - \overline{D}^A \overline{f}_A = 0.\]
Comparing with our calculations of the twistor wave operator, we have proved that the function  $\Xi(f)(x, \eta)$ automatically obeys the twistor wave equation:
\[ (\square (\Xi(f)))(x, \eta) = 0.\]
We have proved:
\begin{itemize} \item The composition of operators $\square \circ \Xi: \mathcal{H}_{-4} \rightarrow \mathcal{H}_{-4}$ vanishes identically.
\end{itemize} 
\eject\noindent
\section{The equation $\Xi \circ \square = 0$}
Suppose that a twistor function $f$ of degree minus four lies in the image of $\square$ acting on $\mathcal{H}_{-2}$, so we have, for some function $g(x, \pi) \in \mathcal{H}_{-2}$ the relations:
\[ \partial_a g = \pi_{A'} g_A + \overline{\pi}_A g_{A'}, \]
\[ (H + 3)g_A = 0, \hspace{10pt} (H + 3) g_{A'} = 0, \]
\[ i(\partial^{A'} g_{A'} - \partial^A g_A)  = f.\]
Henceforth assume that $ \pi_{A'} \eta^{A'} \ne 0$.  Then we may write $g_{A'} = \eta_{A'} q + \pi_{A'} p$, where $p$ and $q$ are functions of $x^a$ and $\pi_{A'}$, with $(H + 3) q = 0$ and $(H + 4)p = 0$.  Put $r = p + \overline{p}$, so $(H + 4)p = 0$.  Note that:
\[ \partial^{A'} g_{A'} =  \partial^{A'}\eta_{A'} q + \partial^{A'} \pi_{A'} p =   \eta_{A'} \partial^{A'} q + (h + 2) p\]
\[ =   \eta_{A'} \partial^{A'} q + \frac{1}{2}(h - \overline{h}) p.\]
Then we have:
\[ \partial_a g(x, \pi) = \pi_{A'} \overline{\eta}_A \overline{q} + \overline{\pi}_A \eta_{A'} q + \overline{\pi}_A \pi_{A'}r, \]
\[ (H + 3) q = 0, (H+ 4) r = 0, \]
\[ i \overline{\eta}_A\overline{\partial}^A \overline{q} - i \eta_{A'} \partial^{A'}q -\frac{i}{2} (h - \overline{h})r  = f.\]
Here $r = |\pi_{C'} \eta^{C'}|^2 \eta^{A'} \overline{\eta}^A \partial_a g$ and $ \pi^{A'} \partial_a g = - \pi_{C'}  \eta^{C'} \overline{\pi}_A q$.
We show that replacing $f$ by the left-hand side of this equation in the integral for $\Xi(f)$, all terms integrate to zero.  We take the defining integral for $\Xi(f)$ in the form:
\[   \Xi(f) = i \int_{\pi_{E'} \eta^{E'} = 1} f(x^a + s\eta^{A'}\overline{\eta}^A, \pi_{A'} )ds \pi^{C'} d\pi_{C'} \overline{\pi}^C d\pi_C.\] 
First consider the $r$ term.  Note that we have, when $\pi_{E'} \eta^{E'} \ne 0$:
\[ \frac{1}{|\pi_{C'} \eta^{C'} |^2} \eta^{A'} \overline{\eta}^A \partial_a (h - \overline{h}) g  = (h - \overline{h})r .\]
Then we have:
\[ ((h - \overline{h}) r)(x^a + s\eta^{A'} \overline{\eta}^A, \pi_{A'}) = \frac{1}{|\pi_{C'} \eta^{C'} |^2} \eta^{A'} \overline{\eta}^A( \partial_a (h - \overline{h}) g) (x^a + s\eta^{A'} \overline{\eta}^A, \pi_{A'}) \]
\[ = \frac{\partial}{\partial s}  \frac{1}{|\pi_{C'} \eta^{C'} |^2}( (h - \overline{h})) g (x^a + s\eta^{A'} \overline{\eta}^A, \pi_{A'}) .\]
Inserting this term into the integral for $\Xi(f)$, the term involving $r$ integrates out to zero.  
\eject\noindent
It remains to show that the term $- i\eta_{A'} \partial^{A'} q$ integrates to zero.   Since the operator $\eta_{A'} \partial^{A'} $ preserves the quantities $\pi_{C'} \eta^{C'}$ and $\overline{\pi}_C \overline{\eta}^C$, we can assume that $\pi_{C'} \eta^{C'} = 1$. Then $d\pi_{A'} = - \eta_{A'} \pi^{C'} d\pi_{C'}$.    In particular $d\pi_{A'} d\pi_{B'} = 0$ and $d\overline{\pi}_A d\overline{\pi}_B = 0$. Then we have, for $w$ a function of $\pi_{A'}$:
\[d (w  \overline{\pi}^C d\overline{\pi}_C ) = \partial^{A'} w d\pi_{A'} \overline{\pi}^C d\overline{\pi}_C\]
\[ = - (\eta_{A'} \partial^{A'} w) \pi^{C'} d\pi_{C'} \overline{\pi}^C d\overline{\pi}_C.\]
Then we have, when $\pi_{A'} \eta^{A'} = 1$: 
\[  (\eta_{E'} \partial^{E'} q)(x^a + s\eta^A\eta^{A'}, \pi_{A'})  ds \pi^{C'} d\pi_{C'} \overline{\pi}^C d\overline{\pi}_C = d(q(x^a + s\eta^A\eta^{A'}, \pi_{A'}) ds \overline{\pi}^C d\overline{\pi}_C).\]
So this term integrates to zero; then by complex conjugation, the $\overline{q}$ term also integrates to zero.   So $\, \Xi(f) = 0$, as required.  We have proved:
\begin{itemize} \item The composition of operators $ \Xi \circ \square: \mathcal{H}_{-2} \rightarrow \mathcal{H}_{-2}$ vanishes identically.
\end{itemize} 
Summarizing, we have proved:
\begin{itemize} \item $ \square \circ \Xi = 0$ vanishes identically, so the kernel of $\square$ contains the image of $\Xi$.
\item $ \Xi \circ \square = 0$ vanishes identically, so the kernel of $\Xi$ contains the image of $\square$.
\end{itemize} 
\eject\noindent
\section{$\mathbb{O}(4, 4)$ triality using quaternions}
Consider three eight-dimensional metric vector spaces $\mathbb{V}_\alpha$, $\mathbb{V}_\beta$ and $\mathbb{V}_\gamma$, whose elements $\lambda$ are represented as pairs of quaternions $\lambda = (P, Q)$, equipped with an $\mathbb{O}(4, 4)$-metric $\lambda.\lambda = |Q|^2 - |P|^2$.   Typical elements of $\mathbb{V}_\alpha$, $\mathbb{V}_\beta$ and $\mathbb{V}_\gamma$ are denoted $\alpha = (A, B)$, $\beta = (C, D)$ and $\gamma = (X, Y)$, respectively.  We introduce a triality $\tau$, a real trilinear form, and three associated products, given as follows:
\[ 2\tau  = 2(\alpha\beta\gamma)  = 2\alpha.(\beta\gamma) = 2 \alpha.(\gamma\beta) = 2\beta.(\gamma\alpha) = 2\beta.(\alpha\gamma) = 2\gamma.(\alpha\beta) = 2\gamma.(\beta\alpha)\]
\[  = - \overline{A}(CY + X\overline{D}) + \overline{B}(\overline{C} X + YD) -  (\overline{Y}\overline{C} + D\overline{X})A + ( \overline{X} C + \overline{D}\overline{Y})B\]
\[  = - \overline{C}(A\overline{Y} - X\overline{B}) + \overline{D}( \overline{Y}B- \overline{A}X) -  (\overline{Y}\overline{A} - B\overline{X})C + ( \overline{B} Y - \overline{X}A)D\]
\[  = - \overline{X}(AD - CB) + \overline{Y}(B\overline{D}  - \overline{C} A) -  (\overline{D}\overline{A} - \overline{B}\overline{C})X + (D \overline{B}  - \overline{A}C)Y.\]
Note that $\tau = (\alpha\beta\gamma) \in \mathbb{R}$.   Here the three real bilinear products are:
\[  (\beta\gamma) = (\gamma\beta) = (C, D)(X, Y) = (CY + X\overline{D}, \overline{C}X + YD) \in \mathbb{V}_\alpha, \]
\[ (\gamma\alpha) = (\alpha\gamma) = (X, Y)(A, B) = (A\overline{Y} - X\overline{B},  \overline{Y}B- \overline{A}X)  \in \mathbb{V}_\beta , \]
\[  (\alpha\beta) = (\beta\alpha) = (A, B)(C, D) =  (AD - CB,  B\overline{D}  - \overline{C} A)  \in \mathbb{V}_\gamma. \]
Then we have the inner products:
\[ (\beta\gamma).(\beta\gamma) = - |CY + X\overline{D}|^2 + |\overline{C}X + YD|^2 =  (|C|^2 - |D|^2)(|X|^2  - |Y|^2) = (\beta.\beta)(\gamma.\gamma) , \]
\[ (\gamma\alpha).(\gamma\alpha) = - |A\overline{Y} - X\overline{B}|^2 + | \overline{Y}B- \overline{A}X|^2 =  (|A|^2 - |B|^2)(|X|^2 - |Y|^2) = (\gamma.\gamma)(\alpha.\alpha), \]
\[ (\alpha\beta).(\alpha\beta) = - |AD - CB|^2 + |B\overline{D}  - \overline{C} A|^2  =  (|A|^2 - |B|^2)(|C|^2 - |D|^2) = (\alpha.\alpha)(\beta.\beta). \]
Next we have the following relations:
\[ (\alpha(\alpha\beta)) = (AD - CB,  B\overline{D}   - \overline{C} A)(A, B)   \]
\[ = (A( D\overline{B}   - \overline{A} C) - (AD - CB)\overline{B},  ( D\overline{B}   - \overline{A} C)B- \overline{A}(AD - CB)) \]
\[ = - (A\overline{A} - B\overline{B})(C, D) = (\alpha.\alpha) \beta, \]
\[ (\beta(\alpha\beta)) = (C, D)(AD - CB,  B\overline{D}   - \overline{C} A)   \]
\[ = (C(B\overline{D}   - \overline{C} A) + (AD - CB)\overline{D}, \overline{C}(AD - CB) + (B\overline{D}   - \overline{C} A)D) \]
\[ = - (C\overline{C} - D\overline{D})(A, B) =  (\beta.\beta) \alpha, \]
\eject\noindent
\[ (\alpha(\gamma\alpha)) = (A, B) (A\overline{Y} - X\overline{B},  \overline{Y}B- \overline{A}X)  \]
\[ =  (A(\overline{Y}B- \overline{A}X) - (A\overline{Y} - X\overline{B})B,  B(\overline{B}Y- \overline{X}A) - (Y\overline{A} - B\overline{X})A) \]
\[ = - (A\overline{A} - B\overline{B})(X, Y) =  (\alpha.\alpha) \gamma, \]
\[ (\gamma(\gamma\alpha)) =  (A\overline{Y} - X\overline{B},  \overline{Y}B- \overline{A}X) (X, Y)  \]
\[ = ((A\overline{Y} - X\overline{B})Y + X(\overline{B}Y - \overline{X}A), (Y\overline{A} - B\overline{X})X + Y(\overline{Y}B- \overline{A}X)) \]
\[ = - (X\overline{X} - Y\overline{Y})(A, B) =  (\gamma.\gamma) \alpha, \]
\[ (\beta(\beta\gamma)) =   (CY + X\overline{D}, \overline{C}X + YD)(C, D)\]
\[ = ((CY + X\overline{D})D - C(\overline{C}X + YD),  (\overline{C}X + YD)\overline{D}  - \overline{C} (CY + X\overline{D})) \]
\[ = - (C\overline{C} - D\overline{D})(X, Y) =  (\beta.\beta) \gamma, \]
\[ (\gamma(\beta\gamma)) =  (X, Y)(CY + X\overline{D}, \overline{C}X + YD) \]
\[ =  ((CY + X\overline{D})\overline{Y} - X(\overline{X}C + \overline{D}\overline{Y}),  \overline{Y}(\overline{C}X + YD)- \overline{A}(CY + X\overline{D})) \]
\[ = - (X\overline{X} - Y\overline{Y})(C, D) =  (\gamma.\gamma) \beta. \]
By polarizing these relations with respect to the reals, we get a series of identities.  For example, when $\gamma' \in \mathbb{V}_\gamma$, we have:
\[ \gamma(\gamma'\alpha) + \gamma'(\gamma\alpha) = 2(\gamma.\gamma' )\alpha.\]
The other key identities for a triality may be verified directly:
\[ ((\gamma\alpha)(\alpha\beta)) = 2(\alpha\beta\gamma) \alpha - (\alpha.\alpha)(\beta\gamma), \]
\[ ((\beta\gamma)(\gamma\alpha)) = 2(\alpha\beta\gamma) \gamma - (\gamma.\gamma)(\alpha\beta), \]
\[ ((\alpha\beta)(\beta\gamma)) = 2(\alpha\beta\gamma)\beta - (\beta.\beta) (\gamma\alpha).\]
We can also prove these identities as follows; pick any $\alpha' \in \mathbb{V}_\alpha$.   Then we have:
\[ \alpha'.((\gamma\alpha)(\alpha\beta)) = ( \alpha'(\gamma\alpha)).(\alpha\beta) = - ( \alpha(\gamma\alpha')).(\alpha\beta) + 2(\alpha.\alpha')\gamma.(\alpha\beta)\]
\[ = -   \alpha(\alpha\beta)).(\gamma\alpha') +   2(\alpha.\alpha')\gamma.(\alpha\beta)  = \alpha'.( - \alpha.\alpha(\beta\gamma) + 2(\alpha\beta\gamma)).\]
Since this formula holds for all $\alpha'$, and since the inner product is non-degenerate, we infer the relation   $((\gamma\alpha)(\alpha\beta)) = 2(\alpha\beta\gamma) \alpha - (\alpha.\alpha)(\beta\gamma)$.   The other identities are proved similarly.
\eject\noindent
Alternatively we can adopt the following approach, using the fact that the products are generically surjective:  for example, write  $\beta = (\alpha\gamma')$, whenever $\alpha.\alpha \ne 0$ (here $\gamma' = (\alpha\beta)(\alpha.\alpha)^{-1} \in \mathbb{V}_\gamma$).  Then we have:
\[ ((\gamma\alpha)(\alpha\beta)) -  2\alpha(\alpha\beta\gamma) + \alpha.\alpha(\beta\gamma)  = ((\gamma\alpha)(\alpha (\alpha\gamma'))) -  2\alpha(\alpha (\alpha\gamma')).\gamma  + \alpha.\alpha( (\alpha\gamma')\gamma)\]
\[ = ( \alpha.\alpha)( (\gamma'(\gamma\alpha)) + (\gamma (\gamma'\alpha)) -  2(\gamma'.\gamma) \alpha) = 0.\]
Since this relation holds for almost all $\alpha \in \mathbb{V}_\alpha$, by continuity, it holds for all $\alpha$.   Finally given any $\alpha \in \mathbb{V}_\alpha$, there is a natural mapping denoted $\hat{\alpha} : \mathbb{V}_\beta \oplus \mathbb{V}_\gamma \rightarrow \mathbb{V}_\beta \oplus \mathbb{V}_\gamma$ given by the formula:  $\hat{\alpha} (\beta \oplus \gamma) = ((\alpha\gamma)\oplus (\alpha\beta))$.
Then the operators $\{ \hat{\alpha}: \alpha \in \mathbb{V}_\alpha\}$, which depend linearly on $\alpha$, represent the Clifford algebra of $\mathbb{O}(4, 4)$, since we have $\hat{\alpha}^2 = (\alpha.\alpha)I$, where $I$ is the identity operator on $\mathbb{V}_\beta \oplus \mathbb{V}_\gamma$.  In particular the space $\mathbb{V}_\beta \oplus \mathbb{V}_\gamma$ may be regarded as the spin space for $\mathbb{V}_\alpha$.  It then follows from the structure theory of the representations of Clifford algebras that the $\mathbb{O}(4, 4)$-triality we have constructed is unique up to isomorphism.    Of course, we also have that $\mathbb{V}_\gamma \oplus \mathbb{V}_\alpha$ may be considered as the spin space for $\mathbb{V}_\alpha$ and $\mathbb{V}_\alpha \oplus \mathbb{V}_\beta$ may be considered as the spin space for $\mathbb{V}_\gamma$.
\\\\We say that  $\alpha\in \mathbb{V}_\alpha$ and $\beta \in \mathbb{V}_\beta$ are \emph{incident} if and only if 
$(\alpha\beta) = 0$.  Multiplying this condition by $\alpha$, we get $(\alpha.\alpha)\beta = 0$; multiplying instead by $\beta$, we get $(\beta.\beta)\alpha = 0$.  It follows that if $\alpha$ and $\beta$ are incident, then either $\alpha =0$, or $\beta =0$ or both $\alpha$ and $\beta$ are null vectors.  Given $\alpha \ne 0$, with $\alpha.\alpha =0$, the space of all $\beta$ such that $\alpha\beta = 0$ turns out to be four-dimensional.    Note also that if $(\alpha\beta) = (\alpha\beta') = 0$, then we have:
\[ 0 = \beta'(\alpha\beta) + \beta(\alpha\beta') = 2(\beta.\beta') \alpha.\]
So $\beta$ and $\beta'$ are necessarily orthogonal, if $\alpha \ne 0$.  Using our quaternionic formalism, the condition $(A, B)(C, D) =0$ becomes:
\[ AD- CB = B\overline{D} - \overline{C} A  = 0.\]
For $(A, B) $ non-zero and null we have $|A|^2 = |B|^2 \ne 0$.   So we may write $(C, D) = (As, tB)$, for some quaternions $s$ and $t$.   Then we need:
\[ 0 = AtB -AsB, \hspace{10pt} 0  = B\overline{B} \overline{t} - \overline{s} \overline{A} A = |B|^2 (\overline{t} - \overline{s}).\]
The general solution is just $s = t$ and we have $(C, D)$ incident with $(A, B) \ne 0$ if and only if $|A|^2 = |B|^2 \ne 0$ and $(C, D) = (At, tB)$, for some quaternion $t$.  Note that the space of solutions is a four-dimensional completely null subspace of $\mathbb{V}_\beta$,  the maximal possible dimension for such a completely null subspace. 
\eject\noindent
\section{The twistor approach to $\mathbb{O}(4, 4)$-triality}
We represent a twistor $Z^\alpha$ by a pair of quaternions: $Z^\alpha = (C, D)$.   The $\mathbb{O}(4, 4)$-inner product of $Z^\alpha$ with itself is $ - |C|^2 + |D|^2  = Z^\alpha \overline{Z}_\alpha$.  We select a unit imaginary quaternion $i$ and write any quaternion $q$ uniquely as sum $q= q_1 + q_2$, where $q_1i = iq_1$ (equivalently $iq_1i = - q_1$) and $q_2 i = - iq_2$ (equivalently $iq_2i = q_2$); explicitly we have and  $q_1= \frac{1}{2}(q - iqi)$ and $q_2 = \frac{1}{2}(q + iqi)$.   If $j$ and $k$ are unit imaginary quaternions, such that $ij  = - ji = k$ and if $q = t + xi + yj + zk$, with $t, x, y, z$ real numbers, then $iqi = - t - xi + yj + zk$, so $q_1 = t + xi$ and $q_2 = yj + zk = (y + iz)j = j(y - iz)$.  We identify the (commutative) subalgebra of all the quaternions $q$ such that $q = q_1$, equivalently $qi = iq$, with the complex numbers $\mathbb{C}$. Then, for any $q\in \mathbb{C}$ and $Z^\alpha = (C, D)$, we define $(q Z)^\alpha = (Cq, q D) $, making the space of all twistors into a four-dimensional complex vector space.     Write $C = c_0 + jc_1$ and $D = d_0 + d_1 j$, where $c_0, c_1, d_0$ and $d_1$ are complex numbers.  Then we may represent $Z^\alpha $ by the complex four-vector $(c_0, c_1, d_0, d_1)$, with the action of $\mathbb{C}$ just given by left multiplication.   The conjugate (dual) vector is then $\overline{Z}_\alpha = (- \overline{c}_0, - \overline{c}_1,  \overline{d}_0,  \overline{d}_1)$.    Consider the twistor description of the triality transformation $Z^\alpha = (C, D) \rightarrow U^\alpha = (CY + X\overline{D}, \overline{C} X + YD)$, where $(X, Y)$ are given quaternions.    This is real linear in the variables $X$ and $Y$, so to understand this  action it suffices to take the four cases: $Y = x \in \mathbb{C}$, $Y = j$, $Y = k$,  $X = x_0 + jx_1$, where $x_0$ and $x_1 $ are complex numbers. 
\begin{itemize} \item  $Y = x $ is just the transformation $Z^\alpha \rightarrow U^\alpha = x Z^\alpha$.
\item  $Y = j$ maps $Z^\alpha = (c_0, c_1, d_0, d_1)$ to $U^\alpha= (- \overline{c}_1, \overline{c}_0, - \overline{d}_1, \overline{d}_0) = A^{\beta\alpha}\overline{Z}_\beta$, with matrix:
\[ A =\hspace{4pt}   \begin{array}{|rrrr|} 0 &1 & 0 &0\\-1&0&0&0\\0&0&0&-1\\0&0&1&0\end{array}.\]   
(The matrix multiplication here is acts on $\overline{Z}_\beta$ from the left with $\overline{Z}_\beta$ treated as a column matrix).
\item  $Y = k$ maps $Z^\alpha = (c_0, c_1, d_0, d_1)$ to $U^\alpha= (i \overline{c}_1, -i\overline{c}_0, -i \overline{d}_1, i\overline{d}_0) = B^{\beta\alpha}\overline{Z}_\beta$, with matrix:
\[ B =\hspace{4pt}   \begin{array}{|rrrr|} 0 &-i & 0 &0\\i&0&0&0\\0&0&0&-i\\0&0&i&0\end{array}.\]  
\item $X = x_0 + jx_1$  maps $Z^\alpha = (c_0, c_1, d_0, d_1)$ to $U^\alpha = (x_0 \overline{d}_0 + \overline{x}_1 \overline{d}_1, x_1 \overline{d}_0 - \overline{x}_0 \overline{d}_1, x_0 \overline{c}_0 + x_1 \overline{c}_1, \overline{x}_1 \overline{c}_0 + \overline{x}_0 \overline{c}_1) = C^{\beta\alpha}\overline{Z}_\beta$, with matrix:
\[ C =\hspace{4pt}   \begin{array}{|rrrr|} 0 & 0& x_0 &\overline{x}_1\\0&0&x_1&-\overline{x}_0\\- x_0&- x_1&0&0\\- \overline{x}_1&\overline{x}_0&0&0\end{array}.\] 
\end{itemize} 
Combining these transformations the full transformation is of the form:
\[ Z^\alpha \rightarrow U^\alpha = x Z^\alpha + X^{\beta\alpha} \overline{Z}_\beta.\]
Here $X^{\alpha \beta} = pA^{\alpha\beta} + qB^{\alpha\beta} + C^{\alpha\beta} = - X^{\beta\alpha}$, where $Y = x + pj + qk$, with $p$ and $q$ real.  It is easily checked that the skew twistor $X^{\alpha\beta}$ obeys the reality condition:
\[ \overline{X}_{\alpha\beta} = \frac{1}{2} \epsilon_{\alpha \beta \gamma \delta} X^{\gamma \delta}.\]
Here $\epsilon_{\alpha\beta\gamma\delta}$ is completely skew and is chosen such that $\epsilon_{1234} = - 1$.      
Finally the space of all skew twistors $X^{\alpha\beta}$ obeying the reality condition is a six-dimensional vector space over the reals and may be parametrized by the quaternion $X$ together with the part, $pj + qk$, of the quaternion $Y$ that obeys $Yi = - iY$. The matrix for $X^{\alpha\beta}$ is:
\[ X^{\alpha\beta} =\hspace{4pt}   \begin{array}{|cccc|} 0 & p- iq& x_0 &\overline{x}_1\\-p+iq&0&x_1&-\overline{x}_0\\- x_0&- x_1&0&-p- iq\\- \overline{x}_1&\overline{x}_0&p+iq&0\end{array}.\]
We have $\frac{1}{4} X^{\alpha\beta}\overline{X}_{\alpha\beta} + |x|^2  = - |X|^2 + |Y|^2$.  Also note that we have: 
\[ U^\alpha U_\alpha = |x|^2 Z^\alpha \overline{Z}_\alpha +  X^{\beta\alpha} \overline{Z}_\beta  \overline{X}_{\gamma\alpha} Z^\gamma  = \left(\frac{1}{4} X^{\alpha\beta}\overline{X}_{\alpha\beta} + |x|^2 \right)Z^\gamma\overline{Z}_\gamma.\]
\eject\noindent
Note that the $\mathbb{O}(4, 4)$-triality $2\tau$ relating $Z^\alpha = (C, D)$, $(X, Y)$ and $\overline{W}^\alpha = (A, B)$ is then the (doubled) inner product $\overline{W}^\alpha \overline{U}_\alpha + U^\alpha W_\alpha$  of $\overline{W}^\alpha = (A, B)$ with $U^\alpha$, so is  
\[  2\tau = x Z^\alpha W_\alpha + \overline{x} \overline{W}^\alpha \overline{Z}_\alpha + X^{\alpha \beta} \overline{Z}_\alpha W_{\beta}  + \overline{X}_{\alpha\beta} Z^\alpha \overline{W}^\beta.  \]
The three triality products are now:
\[ (Z^\alpha,  (x, X^{\alpha\beta})) = W_\alpha =   \overline{x} \overline{Z}_\alpha + \overline{X}_{\beta\alpha} Z^\beta, \]
\[ (W_\alpha, (x, X^{\alpha\beta}))  =  Z^\alpha = x \overline{W}^\alpha -  X^{\beta\alpha} W_\beta, \]
\[ (Z^\alpha, W_\beta) =     (x, X^{\alpha\beta}) = \left(\overline{W}^\alpha \overline{Z}_\alpha, 2Z^{[\alpha} \overline{W}^{\beta]} +  \epsilon^{\alpha\beta\gamma\delta}\overline{Z}_\gamma W_\delta\right).\]
For the last of these products, note that we have:
\[|x|^2 + \frac{1}{4} X^{\alpha\beta} \overline{X}_{\alpha\beta} =  |\overline{W}^\alpha \overline{Z}_\alpha|^2 + 2Z^{[\alpha} W^{\beta]} \overline{Z}_\alpha \overline{W}_\beta = Z^\alpha\overline{Z}_\alpha \overline{W}^\beta W_\beta. \]
We verify directly the triality product structure:
\[ (Z (Z, (x, X))) = (Z^\alpha, \overline{x} \overline{Z}_\alpha + \overline{X}_{\beta\alpha} Z^\beta) \]
\[ = (x Z^\alpha + X^{\beta\alpha} \overline{Z}_\beta)\overline{Z}_\alpha, 2Z^{[\alpha} (x Z^{\beta]} - X^{\beta]\gamma} \overline{Z}_\gamma) +  \epsilon^{\alpha\beta\gamma\delta}\overline{Z}_{\gamma} (\overline{x} \overline{Z}_\delta - \overline{X}_{\delta\epsilon} Z^\epsilon))\]
\[ = (x Z^\alpha\overline{Z}_\alpha, - 2Z^{[\alpha} X^{\beta]\gamma} \overline{Z}_\gamma -  \epsilon^{\alpha\beta\gamma\delta}\overline{Z}_{\gamma} \overline{X}_{\delta\epsilon} Z^\epsilon) \]
Now we have:
\[ - 2Z^{[\alpha} X^{\beta]\gamma} \overline{Z}_\gamma  - \epsilon^{\alpha\beta\gamma\delta} \overline{Z}_\gamma \overline{X}_{\delta \epsilon} Z^\epsilon  \]
\[  =   X^{\alpha\beta} Z^\gamma\overline{Z}_\gamma -   3Z^{[\alpha} X^{\beta\gamma]} \overline{Z}_\gamma   + \frac{1}{2}\epsilon^{\delta\alpha\beta\gamma} \overline{Z}_\gamma \epsilon_{\delta \epsilon\zeta \eta} X^{\zeta \eta} Z^\epsilon \] 
 \[ =  X^{\alpha\beta} Z^\gamma\overline{Z}_\gamma -   3Z^{[\alpha} X^{\beta\gamma]} \overline{Z}_\gamma   + 3 \overline{Z}_\gamma X^{[\alpha\beta} Z^{\gamma]} =  X^{\alpha\beta} Z^\gamma\overline{Z}_\gamma.\]
So we have $Z(Z, (x, X)) = Z^\alpha\overline{Z}_\alpha (x, X)$, as required.  The other verifications are similar and will be omitted.
\eject\noindent
\section{Triality and the invariant $\Xi$-transform}
We assume that the twistor space is equipped with a fixed alternating tensor, $\epsilon_{\alpha\beta\gamma\delta}$, with complex conjugate $\epsilon^{\alpha\beta \gamma \delta}$, normalized by the relation $\epsilon_{\alpha\beta\gamma\delta}\epsilon^{\alpha\beta\gamma\delta} = 24$.   Explicitly we may take $\epsilon^{AB}_{\hspace{12pt} C'D'} = \epsilon^{AB} \epsilon_{C'D'}$ and any spinor parts of $\epsilon^{\alpha\beta\gamma\delta}$, with three or more unprimed indices, or three or more primed indices, vanish.  Let $Z^\alpha$ be a twistor, $W_\alpha$ a dual twistor and let $X^{\alpha\beta}$ be a skew twistor, which obeys the reality condition $X^{\alpha\beta} = \frac{1}{2} \epsilon^{\alpha\beta\gamma\delta} \overline{X}_{\gamma\delta}$.  If the twistor $X^{\alpha\beta}$ has entries $X^{AB} = u\epsilon^{AB}$, $X^A_{\hspace{6pt} B'} = ix^{A}_{B'}$, $X_{A'B'} = v\epsilon_{A'B'}$ with $u$ and $v$ complex and $x^a$ a complex four-vector, then  $Y^{\alpha\beta} = \frac{1}{2} \epsilon^{\alpha\beta\gamma\delta} \overline{X}_{\gamma\delta}$ has entries $Y^{AB} = \frac{1}{2} \epsilon^{AB} \epsilon_{C'D'} \overline{u} \epsilon^{C'D'} = \overline{u} \epsilon^{AB}$,  $Y^{A'B'} = \frac{1}{2} \epsilon^{A'B'} \epsilon_{CD} \overline{v} \epsilon^{CD} = \overline{v} \epsilon^{A'B'}$ and $Y^A_{\hspace{6pt} B'} = -  \epsilon^{AC} \epsilon_{B'D'} \overline{X_{C'}^{\hspace{6pt} D}} = - i\overline{x}^A_{B'}$.     Then the reality condition boils down to the requirement that $u$ and $v$ be real numbers and that  $x^{a}$ be a real four vector.  Note that this reality condition entails the relation:
\[ 4X^{\alpha\beta} \overline{X}_{\gamma\beta} = \delta^\alpha_\gamma X^{\rho \sigma} \overline{X}_{\rho\sigma}.\]
Also we have $X^{\alpha\beta} \overline{X}_{\alpha\beta} = 2u\overline{v} + 2v \overline{u}  + 2X^{A}_{\hspace{6pt} B'} \overline{X_{A'}^{\hspace{7pt}B}}    = 2(u\overline{v} + v \overline{u} + x^a \overline{x}_a)$.   In particular if $X^{\alpha\beta} $ obeys the reality condition, then $ X^{\alpha\beta} \overline{X}_{\alpha\beta}  = 4uv + 2x^ax_a$ (where now $(u, v, x^a)$ are real).   Note that the signature of the quadratic form $X^{\alpha\beta} \overline{X}_{\alpha\beta}$ is $(4, 8)$, for general $X^{\alpha\beta}$ and $(2, 4)$, when $X^{\alpha\beta}$ obeys the reality condition.
As studied in the previous section, the twistor triality formula is:
\[  \tau = Z^\alpha \overline{W}^\beta \overline{X}_{\alpha\beta} + \overline{Z}_\alpha W_\beta X^{\alpha\beta} + x Z^\alpha W_\alpha + \overline{x} \overline{Z}_\alpha \overline{W}^\alpha.\]
Differentiating $\tau$ with respect to $W_\alpha$, we get the incidence relations:
\[ Z^\alpha \overline{X}_{\alpha\beta}  + x\overline{Z}_\beta  = 0, \]
Multiplying this formula by $4X^{\gamma \beta}$, we get:
\[ 0 =  X^{\rho \sigma} \overline{X}_{\rho\sigma} Z^\gamma -  4x \overline{Z}_\beta  X^{ \beta\gamma}  = Z^\alpha( X^{\rho \sigma} \overline{X}_{\rho\sigma} +  4x\overline{x} ).\]
It quickly follows that either $Z^\alpha = 0$, or $X^{\alpha\beta} = 0$ and $x =0$, or $Z^\alpha \overline{Z}_\alpha =  X^{\alpha \beta}\overline{X}_{\alpha\beta} +  4x\overline{x}  = 0$.  Note that provided $X^{\alpha\beta}$ obeys the reality condition,  the signature of the quadratic form $X^{\alpha \beta}\overline{X}_{\alpha\beta} +  4x\overline{x}  = 2(2uv + x^a x_a + 2x\overline{x})$ is $(4, 4)$, the same as that of the basic twistor spaces.
Differentiating $\tau$ with respect to $x$ and $X^{\alpha\beta}$ we get the incidence relations for the twistors $Z^\alpha$ and $W_\alpha$:
\[ Z^\alpha W_\alpha  = 0, \hspace{10pt}  2Z^{[\alpha} \overline{W}^{\beta]} = - \epsilon^{\alpha\beta\gamma\delta}\overline{Z}_\gamma W_\delta.\]
Note that the incidence condition is invariant under real scalings of the twistors $Z^\alpha$ and $W_\alpha$, but not under complex scalings.
Assume that $Z^\alpha$ is incident with $W^\alpha$, where $W_\alpha \ne 0$ is given and $Z^\alpha$ is variable. If $\overline{W}^\alpha W_\alpha \ne 0$, the only solution is $Z^\alpha =  0$.    When $\overline{W}^\alpha W_\alpha =  0$, the solution space is a real vector space of four dimensions and every solution is null: $Z^\alpha \overline{Z}_\alpha = 0$.   Henceforth we take $W_\alpha$ to be non-zero and null. In (conformally compactified) real Minkowski space-time, since $W_\alpha$ is null, $W_\alpha$ selects a null geodesic.   For convenience, we assume that $W_\alpha$ does not lie on the null cone at infinity.  Also for convenience, we delete from our solution space the solutions $Z^\alpha$, for which $Z^\alpha$ and $\overline{W}^\alpha$ are linearly dependent over the complex numbers. Then $Z^\alpha$ is non-zero and null, so it also determines a (variable) null geodesic.   For convenience, we assume that these null geodesics do not meet at infinity.   Then, since $Z^\alpha W_\alpha = 0$, the null geodesic represented by $Z^\alpha$ meets that of $W_\alpha$ at a (variable, finite) point of the null geodesic of $W_\alpha$.  We may write $W_\alpha = (\eta_A, - i x^{BA'} \eta_B)$, for some fixed real vector $x^a$ and fixed non-zero spinor $\eta_A$.  Then, conjugating, we have $\overline{W}^\alpha = (i x^{B'A} \overline{\eta}_{B'}, \overline{\eta}_{A'})$.   Also the twistor $Z^\alpha$ may be written: $Z^\alpha = (i(x^{AB'} + s\eta^A \overline{\eta}^{B'})\pi_{B'}, \pi_{A'})$, where $s$ is a real variable and $\pi_{A'}$ is a non-zero variable spinor subject to the condition that $\pi_{A'} \overline{\eta}^{A'}$ be pure imaginary and non-zero.  Write $\pi_{A'} \overline{\eta}^{A'} = i u $, where $u$ is real and non-zero.  Note that the condition $u \ne 0$ guarantees that $Z^\alpha$ and $\overline{W}^\alpha$ are linearly independent over the complex numbers and that they meet at a finite point, as required.  Then we have:
\[ dZ^\alpha = - u ds (\eta^A, 0) + (i(x^{AB'} + s\eta^A \overline{\eta}^{B'})d\pi_{B'}, d\pi_{A'}).\]
Now we may write $d\pi_{A'} =  \alpha \pi_{A'} + \beta \overline{\eta}_{A'}$, where the one-form $\beta$ is complex whereas the one-form $\alpha$ is real.  Note that $\pi^{A'} d\pi_{A'} = - iu\beta$.  Also define the fixed auxiliary twistor $U^\alpha = (\eta^A, 0)$ (so $\overline{U}_\alpha = (0, \overline{\eta}^{A'})$.  Note that the twistors $U^\alpha$, $Z^\alpha$ and $\overline{W}^\alpha$ are linearly independent.  Also we have $Z^\alpha \overline{U}_\alpha = iu$.  Then we have:
\[ dZ^\alpha = -  u ds U^\alpha + \alpha Z^\alpha + \beta \overline{W}^\alpha.\]
In particular we have for the contact form:
\[ i \overline{Z}_\alpha dZ^\alpha =  u^2ds.\]
\eject\noindent
Next let $\theta^\alpha$ be an auxiliary twistor-valued (constant) Grassman variable that anti-commutes with itself, with the exterior derivative operator and with its conjugate, $\overline{\theta}_\alpha$.  Consider the four-form:
\[ \omega_4 = \frac{1}{24}  (\theta^\alpha d\overline{Z}_\alpha + \overline{\theta}_\alpha dZ^\alpha)^4.\]
Put $U = U^\alpha \overline{\theta}_\alpha + \overline{U}_\alpha \theta^\alpha$, $Z = Z^\alpha \overline{\theta}_\alpha + \overline{Z}_\alpha \theta^\alpha$ and $W = W_\alpha \theta^\alpha$.  Then we have:
\[ \omega_4 = \frac{1}{24}  ( -u U ds + Z \alpha + W \overline{\beta} + \overline{W} \beta)^4  =   - u U ds  Z \alpha W \overline{\beta} \overline{W} \beta =    (UZW\overline{W}) uds \alpha \beta  \overline{\beta} .\]
The real vector field $H = Z^\alpha \partial_\alpha + \overline{Z}_\alpha \overline{\partial}^\alpha$ is tangent to the relations obeyed by $Z^\alpha$.   Contracting $dZ^\alpha$ with $H$ gives the relation:
\[ \iota_{H} dZ^\alpha = Z^\alpha = \iota_{H} (- U^\alpha u ds + \alpha Z^\alpha + \beta  \overline{W}^\alpha).\]
It follows that $\iota_{H} (ds) = \iota_{H} (\beta) = 0$ and $\iota_{H}(\alpha) = 1$.  Define $\omega_3 = - \iota_H \omega_4$.  Then we have:
\[\omega_3 = - \iota_H \omega_4 = \frac{1}{6}(\theta^\alpha \overline{Z}_\alpha + \overline{\theta}_\alpha Z^\alpha)  (\theta^\beta d\overline{Z}_\beta + \overline{\theta}_\beta dZ^\beta)^3  =    UZW\overline{W} u ds \beta  \overline{\beta} .\]
Note that $\omega_3$ is a projective form of weight four: under the scaling $Z^\alpha \rightarrow t(Z) Z^\alpha$, where $t(Z)$ is a non-zero function of $Z^\alpha$, we have $\omega_3 \rightarrow t(Z)^4 \omega_3$.  Next we work out the Grassman element $UZW\overline{W}$.  Put $\phi^A = \theta^A - ix^a \theta_{A'}$ and $\upsilon =  \overline{U}_\alpha \theta^\alpha =  \overline{\eta}^{A'} \theta_{A'} $.     Then we have:
\[ W = W_\alpha \theta^\alpha = \eta_A \theta^A - i x^a\eta_A \theta_{A'} = \eta_A \phi^A,\]
\[ U =  \upsilon + \overline{\upsilon}, \]
\[ \overline{Z}_\alpha \theta^\alpha = - i(x^a + s\eta^A\overline{\eta}^{A'})\overline{ \pi}_{A} \theta_{A'}  + \overline{\pi}_A \theta^A = \overline{\pi}_A \phi^A - is \eta^A\overline{ \pi}_{A}\upsilon,  \]
\[ Z = \overline{\pi}_A \phi^A + \overline{\pi}_{A'} \overline{\phi}^{A'} - is \eta^A\overline{ \pi}_{A} U, \]
\[ UZ W\overline{W} = (\upsilon + \overline{\upsilon})( \overline{\pi}_A \phi^A + \overline{\pi}_{A'} \overline{\phi}^{A'} ) ( \eta_B \phi^B  \overline{\eta}_{B'} \overline{\phi}^{B'})\]
\[ = - \frac{i}{2} u (\upsilon + \overline{\upsilon})( \phi_B \overline{\eta}_{B'} - \eta_B \overline{\phi}_{B'})\phi^B \overline{\phi}^{B'}  =  i u \Psi, \]
\[ \Psi = - \frac{1}{2} (\upsilon + \overline{\upsilon})( \overline{\eta}_{B'}\phi_B - \eta_B \overline{\phi}_{B'})\phi^B \overline{\phi}^{B'} = \overline{\Psi}.\]
\eject\noindent
We next show that the quantity $\Psi$ may be neatly expressed as a quadratic in the twistor $W_\alpha$.
Note that:
\[ \theta^\alpha \overline{\theta}_\alpha W \overline{W} = (\theta^A \overline{\theta}_A + \theta_{A'} \overline{\theta}^{A'} ) W \overline{W} \]
\[ =  ((\phi^A + ix^a \theta_{A'}) \overline{\theta}_A + \theta_{A'}( \overline{\phi}^{A'} - i x^a \theta_A)) W \overline{W}\]
\[ =  (\phi^A \overline{\theta}_A + \theta_{A'}\overline{\phi}^{A'}  ) \eta_B \phi^B  \overline{\eta}_{B'} \overline{\phi}^{B'}\]
\[  = - \frac{1}{2} ( \overline{\upsilon}  \phi_B \overline{\eta}_{B'} - \upsilon \overline{\phi}_{B'} \eta_B) \phi^B \overline{\phi}^{B'}.\]
Also put $\theta^\alpha \theta^\beta \theta^\gamma = \epsilon^{\alpha\beta\gamma\delta} \Sigma_\delta$, so $\Sigma_\alpha = - \frac{1}{6} \epsilon_{ \alpha\beta\gamma\delta}  \theta^\beta \theta^\gamma \theta^\delta$.  We have:
\[ \Sigma_A = - \frac{1}{2} \epsilon_{AB}\epsilon^{C'D'} \theta^B \theta_{C'} \theta_{D'}  =  \frac{1}{2}\phi_A \theta_{C'} \theta^{C'},   \]
\[ \Sigma^{A'} = - \frac{1}{2} \epsilon^{A'B'}\epsilon_{CD}  \theta_{B'} \theta^C \theta^D  = - \frac{1}{2}   \theta^{A'} \theta_C \theta^C\]
\[ = - \frac{1}{2}   \theta^{A'} (\phi_C + 2i x_{C}^{D'}\theta_{D'})  \phi^C  = - \frac{1}{2}   \theta^{A'} \phi_C  \phi^C - \frac{i}{2}   x^a  \phi_A\theta_{C'} \theta^{C'}\]
\[ = - \frac{1}{2}   \theta^{A'} \phi_C  \phi^C - ix^a \Sigma_A, \]
\[ \overline{W}^\alpha \Sigma_\alpha = \overline{\eta}_{A'} \Sigma^{A'} + ix^a \overline{\eta}_{A'} \Sigma_{A} =   - \frac{1}{2}  \overline{\eta}_{A'}   \theta^{A'} \phi_C  \phi^C =  \frac{1}{2} \upsilon \phi_C  \phi^C, \]
\[   \overline{W}^\alpha \overline{W}^\beta  \Sigma_\alpha \overline{\theta}_\beta =  \frac{1}{2}   \upsilon  \overline{\eta}_{B'}  \overline{\phi}^{B'} \phi_C  \phi^C. \] 
Expanding out $\Psi$, we get:
\[ \Psi = - \frac{1}{2}\upsilon \overline{\eta}_{B'} \overline{\phi}^{B'}\phi_B\phi^B  - \frac{1}{2} \upsilon  \eta_B\phi^B  \overline{\phi}_{B'}\overline{\phi}^{B'} + c.c.\]
\[ = \overline{W}^\alpha \overline{W}^\beta   \overline{\theta}_\alpha \Sigma_\beta    + \theta^\alpha \overline{\theta}_\alpha W \overline{W} +  W_\alpha W_\beta \theta^\alpha \overline{\Sigma}^\beta.\]
So now we have a concrete expresssion for the three-form $\omega_3$:
\[  \omega_3 = UZW\overline{W} u ds \beta  \overline{\beta} \]
\[  =  (\overline{W}^\alpha \overline{W}^\beta   \overline{\theta}_\alpha \Sigma_\beta    + \theta^\alpha \overline{\theta}_\alpha W \overline{W} +  W_\alpha W_\beta \theta^\alpha \overline{\Sigma}^\beta)i u^2 ds \beta \overline{\beta}\]
\[ =  i  (\overline{W}^\alpha \overline{W}^\beta   \overline{\theta}_\alpha \Sigma_\beta    + \theta^\alpha \overline{\theta}_\alpha W \overline{W} +  W_\alpha W_\beta \theta^\alpha \overline{\Sigma}^\beta) ds \pi^{C'} \overline{\pi}^C d\pi_{C'} d\overline{\pi}_C.\]
Let $f(Z)$ be a given twistor function of degree minus four: $f(tZ) = t^{-4} f(Z)$, for any real non-zero real number $t$. The invariant $\Xi$-transform of $f(Z)$ is by definition:
\[ \Xi(f, \theta, W) = \int_{Z \hspace{3pt} \textrm{incident with}\hspace{3pt} W}  f(Z) \omega_3.\]
Here $W_\alpha$ is any non-zero null twistor and $f(Z)$ is defined on $\mathbb{N}'$.    The integral is taken over a three-sphere and always converges, with the result varying smoothly with $W$.  Comparing with our original definition of the transform $\Xi(f)$, we find that we have proved the fundamental fact:
\begin{itemize} \item $\displaystyle{\Xi(f, \theta, W)  = \Psi \Xi(f) =  (\overline{W}^\alpha \overline{W}^\beta   \overline{\theta}_\alpha \Sigma_\beta    +W_\alpha \overline{W}^\beta \theta^\gamma \overline{\theta}_\gamma \theta^\alpha \overline{\theta}_\beta +  W_\alpha W_\beta \theta^\alpha \overline{\Sigma}^\beta) \Xi(f)}$.
\end{itemize} 
Note that since the left-hand side of this equation is plainly invariant under the scaling $W_\alpha \rightarrow  tW_\alpha$, where $t$ is any real non-zero number, so we see immediately that $\Xi(f)$ must be of degree minus two in $W_\alpha$, which was previously established by direct calculation. 
\eject\noindent  
\section{The reduction of the invariant $\Xi$-transform to the $\mathbb{SU}(2, \mathbb{C})$ transform}
We consider the invariant $\Xi$-transform:
\[  \Xi(f)(W, \theta) = \int_{Z \hspace{3pt} \textrm{incident with} \hspace{3pt} W} f(Z) Z.\theta (dZ.\theta)^3.\]
Here $\theta$ is a vector-valued Grassman variable.  We represent $Z$ by the quaternion pair $(C, D)$ and $W$ by the quaternion pair $(A, B)$, such that the incidence condition of $Z$ with $W$ reads $(C, D) = (At, tB)$, for some quaternion $t$, where $|C| = |D| \ne 0$ and $|A| = |B| \ne 0$.  Also $f(Z)$ is homogeneous of degree minus four, so the integral is both projectively invariant and invariant under the non-zero real scalings of $W$, so we may assume without loss of generality that $|A|^2 = |B|^2 = |C|^2 = |D|^2 = |t|^2 = 1$.  Write $\theta = (\alpha, \beta) $, where $\alpha$ and $\beta$ are quaternion-valued Grassman variables, such that $Z.\theta = \overline{\alpha}C + D \overline{\beta} + \overline{C} \alpha + \overline{\beta} \overline{D}$ and $dZ.\theta =  - \overline{\alpha}dC + dD \overline{\beta} + d\overline{C} \alpha - \overline{\beta} d\overline{D}$.   Substituting for $Z$ in terms of $W$, we have:
\[ Z.\theta = \overline{\alpha}At + tB \overline{\beta} + \overline{t} \overline{A} \alpha + \beta \overline{B} \overline{t},\]
\[ dZ.\theta =  - \overline{\alpha}Adt + dt B \overline{\beta} + d\overline{t}\overline{A} \alpha - \beta  \overline{B} d\overline{t}.\]
Next write $\alpha = A\gamma$ and $\beta = \delta B$, where $\gamma = \overline{A} \alpha$ and $\delta = \beta \overline{B}$ are Grassman variables.   \\Then we have:
\[ Z.\theta =  \overline{\gamma}t + t \overline{\delta} + \overline{t}\gamma + \delta  \overline{t},\]
\[ dZ.\theta =  - \overline{\gamma}dt + dt \overline{\delta} + d\overline{t} \gamma - \delta d\overline{t}.\]
Using $\mathbb{R}^4$ indices, we may write:
\[ Z.\theta = \epsilon_at^a,  \hspace{10pt} dZ.\theta = - \epsilon_bdt^b,  \hspace{10pt} \epsilon_a= 2(\gamma_a + \delta_a). \]
We then have:
\[ Z.\theta (dZ.\theta)^3 =  \epsilon_a \epsilon_b  \epsilon_c \epsilon_d  t^adt^bdt^cdt^d \]
\[ = \frac{1}{6} \Omega_W(\theta) \epsilon_{abcd} t^a dt^b dt^c dt^d, \]
\[ \Omega_W(\theta) = \frac{1}{4} \epsilon^{abcd} \epsilon_a  \epsilon_b  \epsilon_c  \epsilon_d.\] 
So after factoring out the form $\Omega_W (\theta)$, and writing $p$ for $t$, $g$ for $A$ and $h$ for $B$, the transformation becomes just:
\[ \Xi(f)(g, h) = \int_\mathbb{G} f(gp, ph) \omega_p.\]
This exactly agrees with our original $\Xi$-transform on the Lie group $\mathbb{SU}(2, \mathbb{C}) \times \mathbb{SU}(2, \mathbb{C}$, completing the demonstration that all three transformations are equivalent.\\\\
Note that written invariantly, without the use of the Grassman variable $\theta$, we have the following expression for the transform:
\[  \Xi(f)(W) \sigma^{abcd}_{\alpha\beta} W^\alpha W^\beta  = \int_{Z \hspace{4pt} \textrm{incident with} \hspace{4pt} W}  f(Z) Z^{[a} dZ^b dZ^c dZ^{d]}.\]
Here, as in the last section, the quantity $\sigma^{abcd}_{\alpha\beta}$ gives a natural isomorphism from trace-free symmetric tensors $X^{\alpha\beta}$ in eight dimensions to real self-dual four forms $X^{abcd}$ in eight dimensions, each space being thirty-five dimensional.

\end{document}